\documentclass[sigconf]{acmart}

\newcommand {\codename}{SlideAudit\xspace}
\AtBeginDocument{%
  }

\copyrightyear{2025}
\acmYear{2025}
\setcopyright{cc}
\setcctype{by}
\acmConference[UIST '25]{The 38th Annual ACM Symposium on User Interface
Software and Technology}{September 28-October 1, 2025}{Busan, Republic of
Korea}
\acmBooktitle{The 38th Annual ACM Symposium on User Interface Software and
Technology (UIST '25), September 28-October 1, 2025, Busan, Republic of
Korea}\acmDOI{10.1145/3746059.3747736}
\acmISBN{979-8-4007-2037-6/2025/09}

\usepackage{listings}
\usepackage{xspace}
\usepackage{multirow}

\usepackage{xparse}

\usepackage{alltt}

\NewDocumentCommand{\codeword}{v}{%
\texttt{\textcolor{black}{#1}}%
}

\usepackage{xcolor}

\begin{document}

\newcommand{\jerry}[1]{{\textcolor{black}{#1}}}
\title[\codename: A Dataset and Taxonomy for Automated Evaluation of Presentation Slides]{\codename: A Dataset and Taxonomy for \\ Automated Evaluation of Presentation Slides}

\author{Zhuohao (Jerry) Zhang}
\affiliation{%
  \institution{University of Washington}
  \city{Seattle}
  \state{Washington}
  \country{United States}
}
\email{zhuohao@uw.edu}

\author{Ruiqi Chen}
\affiliation{%
  \institution{University of Washington}
  \city{Seattle}
  \state{Washington}
  \country{United States}
}
\email{ruiqich@uw.edu}

\author{Mingyuan Zhong}
\affiliation{%
  \institution{University of Washington}
  \city{Seattle}
  \state{Washington}
  \country{United States}
}
\email{myzhong@cs.washington.edu}

\author{Jacob O. Wobbrock}
\orcid{0000-0003-3675-5491}
\affiliation{%
  \institution{University of Washington}
  \city{Seattle}
  \state{Washington}
  \country{United States}
}
\email{wobbrock@uw.edu}

\renewcommand{\shortauthors}{Zhang et al.}

\begin{abstract}

Automated evaluation of specific graphic designs like presentation slides is an open problem. We present \textit{\codename}, a dataset for automated slide evaluation. 
We collaborated with design experts to develop a thorough taxonomy of slide design flaws. Our dataset comprises 2400 slides collected and synthesized from multiple sources, including a subset intentionally modified with specific design problems. We then fully annotated them using our taxonomy through strictly trained crowdsourcing from \textit{Prolific}. To evaluate whether AI is capable of identifying design flaws, we compared multiple large language models under different prompting strategies, and with an existing design critique pipeline. We show that AI models struggle to accurately identify slide design flaws, with F1 scores ranging from 0.331 to 0.655. Notably, prompting techniques leveraging our taxonomy achieved the highest performance. We further conducted a remediation study to assess AI's potential for improving slides. Among 82.0\% of slides that showed significant improvement, 87.8\% of them were improved more with our taxonomy, further demonstrating its utility.

\end{abstract}


\maketitle

\section{Introduction}

Creating visually effective graphics such as presentation slides remains a complex design task. It demands careful attention to layout structure, visual hierarchy, color usage, and communicative clarity—all of which significantly influence how information is perceived by an audience. While recent advances in AI have enabled automated slide generation \cite{bandyopadhyay2024enhancingpresentationslidegeneration, gemini, wang2024outlinesparkignitingaipoweredpresentation, aggarwal2025passpresentationautomationslide}, identifying and improving design issues in existing slides remains a challenging and underexplored problem. In practice, users often work with existing templates or partially completed slides, iteratively refining them rather than generating complete decks from scratch. Yet, diagnosing and correcting design flaws (e.g., issues affecting clarity, aesthetics, or communication) continues to rely heavily on human judgment, with limited automated support available.

To investigate this open challenge, we must first understand the scope of potential design problems in presentation slides. While extensive research exists on design critique and automated design assessments \cite{luther2015structuring, duanUICritEnhancingAutomated2024, alabood2023systematic}, there remains a critical gap in the systematic categorization of design flaws, specifically in presentation slides. We developed \textit{\codename}, a dataset of slides annotated according to a rigorously defined taxonomy of slide design flaws. We first conducted a formative design study with 17 experts to iteratively derive and validate the taxonomy of design flaws. We then assembled an expansive dataset of 2400 slides from multiple sources, including an existing slide corpus, publicly-shared Google Slides presentations, and slide decks generated by AI systems. To model realistic variation in slide quality, we synthetically introduced controlled manipulations related to object positioning, layout patterns, typography, and color distributions. In total, we used a combination of 600 originally collected slides and 1800 synthesized slides. Using strictly qualified crowdworkers recruited through \textit{Prolific},\footnote{Prolific is a crowdsourcing platform (https://www.prolific.com/)} we annotated each slide according to our proposed taxonomy.

We conducted three evaluations to examine both our dataset's practical utility and the feasibility of leveraging large language models (LLMs) to automatically detect and remediate slide flaws. First, we evaluated LLM-based detection of design flaws using four prompting strategies~\cite{bsharat2023principled}: (1) baseline with no additional guidance, (2) high-level categorical knowledge augmentation, (3) comprehensive taxonomy descriptions paired with examples, and (4) computationally augmented prompts. Second, we compared \codename with prior design critique frameworks, CrowdCrit \cite{luther2015structuring} and UICrit \cite{duanUICritEnhancingAutomated2024}. We applied design principles derived in CrowdCrit and generated design critiques to the same evaluation process to show differences between CrowdCrit- and SlideAudit-based LLM variants. We also directly adapted and evaluated UICrit's evaluation pipeline with SlideAudit data, applying its few-shot and visual prompting approaches to identify slide flaws, allowing a different type of comparison with our method. Third, we conducted a follow-up participant study examining the effectiveness of LLM-generated remediation plans, comparing remediation quality when taxonomy-informed diagnostics were provided versus absent.

Our results show that LLMs alone struggle to consistently identify slide design flaws, achieving varied performance (F1: 0.476 -- 0.655). However, explicitly including our taxonomy improved model accuracy in all tests. In the remediation evaluation with human raters, 82.0\% of slides showed significant improvement from at least one generated plan. Among these improved slides, the taxonomy-informed approach generated better remediation plans in 87.8\% of cases when compared to the baseline. Notably, providing the taxonomy also enabled the correct identification of defect-free slides, reducing unproductive remediation.

In summary, this work contributes:
\begin{itemize}
\item A rigorously-developed taxonomy of slide design flaws derived from iterative expert collaboration;
\item \codename, a richly-annotated dataset of 2400 slides systematically annotated with flaw information;
\item Evaluations demonstrating the benefit of incorporating the taxonomy into AI-driven slide critique and remediation.
\end{itemize}

\section{Related Work}

Our research builds on related work in automated design critiques and evaluations, datasets related to visual understanding, and creativity support for blind and low-vision users. We address each of these areas in turn, below.

\subsection{Design Evaluations}

Traditional graphic design evaluations typically rely on expert critique informed by theories and principles of visual perception. Influential frameworks such as the Gestalt principles \cite{wertheimer2017untersuchungen} describe how people perceptually group visual elements according to factors like proximity, similarity, continuity, and closure. Violations to the principles would affect the presentation qualities of graphic designs. Principles from visual hierarchy and perception~\cite{Gordon2020,IxDF2025,lidwell2010universal,ware2019information,nielsen1994enhancing} further guide critiques by helping evaluators assess clarity, effective information communication, and aesthetic composition. In practice, designers frequently apply these frameworks through structured heuristic evaluations \cite{nielsen1990heuristic, nielsen10usability, nielsen1992finding}, expert reviews \cite{Harley2018}, crowdsourcing \cite{luther2015structuring, xu2014voyant, yuan2016almost}, and design checklists informed by established guidelines \cite{Friedman2022,Nova2023,jambor2024zero}. Despite offering valuable qualitative insights and actionable suggestions, these conventional approaches rely heavily on evaluator expertise, iterative manual effort, and visual discernment. For blind professionals needing to produce visually effective materials, these traditional evaluation approaches pose substantial accessibility challenges, typically requiring them to rely extensively on sighted colleagues or peers for critical feedback and improvement.

\subsection{Automated Visual Design Evaluation}

Automated visual evaluation methods emerged predominantly from rule-based heuristics and predictive models trained on manually labeled domain-specific datasets~\cite{lee2020guicomp, oulasvirta2018aalto, duan2020optimizing, fosco2020predicting, wu2020predicting}. However, the applicability of these earlier approaches was limited due to narrow evaluation scopes and poor cross-domain generalizability. Recent research thus introduced richer datasets containing diverse visual attributes, layouts, and human-usage patterns for user interface (UI) and slide design domains~\cite{deka_rico_uist17, wu2023webui, fosco2020predicting, jiang2023ueyes, leiva2020understanding, wu2023never}. For slides specifically, the SpaSe~\cite{haurilet2019spase} and WiSE~\cite{haurilet2019wise} datasets provided manually annotated segmentation masks from presentation slides. Other datasets focused on annotating slide elements for tasks, such as visual question-answering or retrieval~\cite{tanaka2023slidevqa}; moreover, multimodal slide datasets provided structural element labeling~\cite{jobin2024semantic,kim2022fitvid, lee2023lecture}. Peng et al.'s DreamStruct~\cite{peng2024dreamstruct} provides synthetic structured graphic inputs, facilitating deeper visual understanding tasks for presentations and UIs. Collectively, these datasets contributed foundational resources for automated methods capable of general-purpose visual understanding and downstream evaluation tasks. Building on these prior dataset contributions, we introduce a structured approach leveraging LLMs called \textit{\codename} to automatically identify and categorize visual design flaws specifically from raw slide images.

Recent advances in large language models LLMs have opened novel possibilities for design and evaluation tasks, moving beyond manually engineered heuristics \cite{jiang2025iluvui, wu2023never, schoop2022predicting, zhang2021screen}. Duan et al. \cite{duan2024generating, duanUICritEnhancingAutomated2024}, for instance, leveraged LLMs to produce flexible UI critiques and feedback enabled by few-shot and visual prompting techniques. Similarly, Wu et al. developed UIClip \cite{wu2024uiclip}, a multimodal vision-language approach quantifying UI design quality to enhance LLM-assisted UI code generation. Extending these successes into broader design critiques, researchers have demonstrated LLM effectiveness across other complex tasks, such as accessibility assessment~\cite{taeb24axnav}, programming error diagnostics~\cite{zhangRepairingBugsPython2022}, academic reviewing~\cite{liangCanLargeLanguage2023a}, and reasoning about agentic action's outcomes~\cite{zhang2025interaction}. Such work indicates strong capabilities of LLMs in contextual task comprehension, structured reasoning, and actionable critique generation.

Motivated by these advances, we explore the feasibility of employing LLMs for automatically evaluating visual design flaws specifically on presentation slides, which has not been explored in prior literature. Specifically, \codename investigates whether systematically categorizing design flaw types improves LLM accuracy in identifying and remedying flaws in raw slide graphics.

\begin{figure*}[h]
    \centering
    \includegraphics[width=\textwidth]{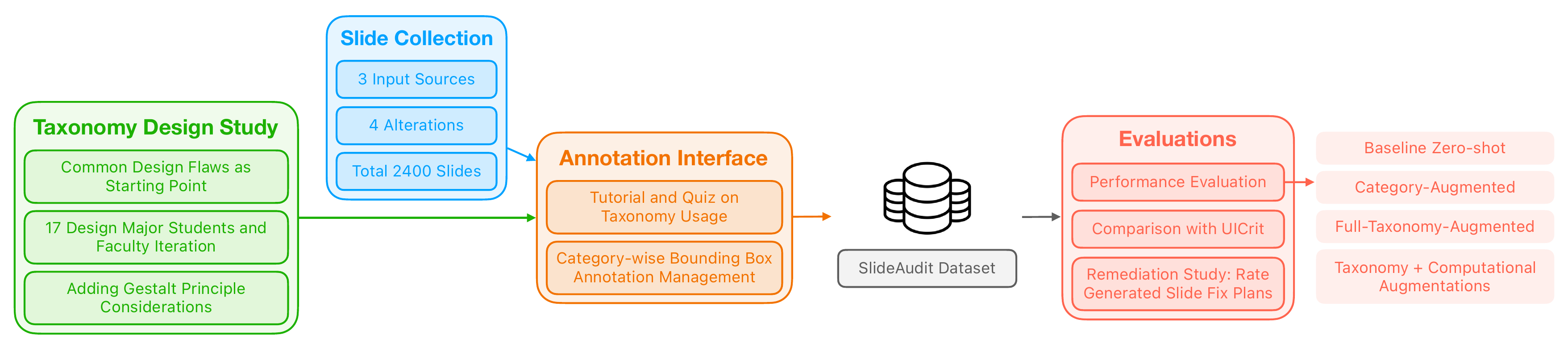}
    \caption{The \codename pipeline, covering taxonomy design, slide dataset creation, annotation, and evaluation stages.}
    \label{fig:slideaudit_pipeline}
\end{figure*}

\section{Taxonomy of Slide Design Flaws}

Although some slide design issues such as low contrast or misalignment may be readily apparent, systematically categorizing these flaws into clear and mutually exclusive groups remains challenging. Existing approaches often document surface-level issues without fully addressing underlying perceptual principles. To enable structured and theoretically grounded assessment of visual slide quality, we iteratively conducted a development study informed by literature, designers, and an expert in visual communication design.

\subsection{Review of Prior Taxonomies}
Prior work has established several graphic design guidelines \cite{o2015designscape, o2014learning} and taxonomies for visual design evaluation \cite{luther2015structuring, duanUICritEnhancingAutomated2024}, including CrowdCrit~\cite{luther2015structuring} for general graphic design and UICrit~\cite{duanUICritEnhancingAutomated2024} for user interface evaluation. DesignChecker~\cite{huh2024designchecker} also employed a structured set of issue types for web design assessment. These frameworks provide valuable foundations for understanding design quality across various media.

However, presentation slides introduce distinct design considerations that are not fully addressed by existing taxonomies. As Dourish and Button~\cite{dourish1998technomethodology} noted, design practices are locally situated, and the domain of slide design reflects this: slides operate as sequential, time-based narratives that must maintain coherence across slides, support temporal unfolding through transitions and animations, and balance live presentation needs with asynchronous readability. These characteristics raise challenges that differ meaningfully from static graphic design or interactive user interfaces, requiring dedicated criteria for structured evaluation.

Additionally, while prior frameworks such as CrowdCrit offer rich insights (e.g., 70 open-ended critique statements grouped into approximately 33 themes), these were not intended as a finalized or mutually exclusive taxonomy. Instead, they served to scaffold human evaluators through flexible guidance. Existing frameworks also tend to emphasize breadth and interpretive richness, rather than compactness or automation. As such, there remains a need for a more systematically defined and domain-specific taxonomy that reflects slide-specific concerns while offering a clear structure for both human and automated evaluation. Building on these insights, we aim to develop a slide-specific taxonomy that is compact, mutually exclusive, and grounded in design theory---capable of addressing the unique demands of presentation slides and supporting both human and automated evaluation.

\subsection{Method}
We conducted a multi-phase study (Figure~\ref{fig:slideaudit_pipeline}) incorporating expert evaluation and critical reflection, as described below.

\subsubsection{Participants}
We recruited 16 students from different design majors to assess visual design flaws in slides. The participant group consisted of 7 senior undergraduate students and 9 students earning their Master's degrees. Their majors included design and visual communication, with specific program names omitted to maintain anonymity. All participants had extensive experience creating and evaluating presentation slides. Pre-study responses indicated that participants regularly engage in slide design critique across diverse contexts, including academic peer reviews, professional workplace settings, collaborative projects, and informal assistance to colleagues. Their critique experience ranged from improving visual hierarchy and readability to developing brand-consistent slide templates and refining presentation materials for job interviews and client pitches.

Additionally, we engaged one visual communication design faculty member with over 30 years of experience in design and visual communication. The faculty member contributed to the study, as described below, after reviewing student responses.

\subsubsection{Apparatus}
The study was guided by a detailed document that helped participants explore and refine the design flaw taxonomy. The document was structured to progressively introduce participants to the categories, starting with broad groupings and narrowing down to more specific subcategories. This approach helped participants build a conceptual understanding of the taxonomy before applying it to real-world examples. Additionally, participants were encouraged to critically evaluate the taxonomy, suggesting new categories and identifying ambiguities.

\subsubsection{Procedure}
The study followed a multi-stage procedure designed to engage participants with the classification system while collecting their feedback. The stages were:

\begin{itemize}

\item \textit{Progressive Exploration of Categories}: Participants were introduced to the design flaw categories through a structured, hierarchical presentation. They began with general categories and moved to more specific subcategories. This gradual unfolding helped participants grasp the overall structure of the taxonomy and understand its distinctions.

\item \textit{Application to Concrete Examples}: After becoming familiar with the taxonomy, participants analyzed slides with pre-annotated flaws. This hands-on phase helped solidify their understanding of how to recognize and categorize common design issues.

\item \textit{Critical Challenge Phase}: As participants gained confidence with the taxonomy framework, they engaged in adversarial thinking. They were tasked with identifying or creating slides containing subtle design issues that might challenge automated detection systems. This phase pushed participants to consider nuanced design problems beyond obvious violations, including ``cleverly hidden flaws'' and potential ``false-positive traps.''

\item \textit{Reflective Evaluation}: Finally, participants reflected on the taxonomy, assessing its strengths and weaknesses. They provided feedback on missing categories, unclear definitions, and areas for improvement.

\item \textit{Expert Analysis}: The faculty expert participated in a synchronous session to review the taxonomy and student feedback. In addition to suggesting refinements, the expert guided a shift in framing from identifying surface-level design issues to incorporating audience-centered and perceptual principles rooted in design theory.
\end{itemize}

\subsubsection{Analysis}
We conducted a thematic analysis~\cite{clarke2017thematic} of participant responses. Two researchers independently coded the qualitative data using an initial codebook derived from preliminary taxonomy categories. Inter-rater reliability was quantified using Cohen’s kappa, with a result of $\kappa = 0.66$ (indicating substantial agreement between coders). Throughout two iterative coding cycles, the researchers discussed and resolved definitional ambiguities and any coding disagreements until consensus (full agreement between coders) was reached on all codes. The final analysis consolidated the agreed-upon codes into higher-level themes aligned with the study's goals, directly informing the refined taxonomy structure.

\begin{figure*}[htbp]
    \centering
    \includegraphics[width=\textwidth]{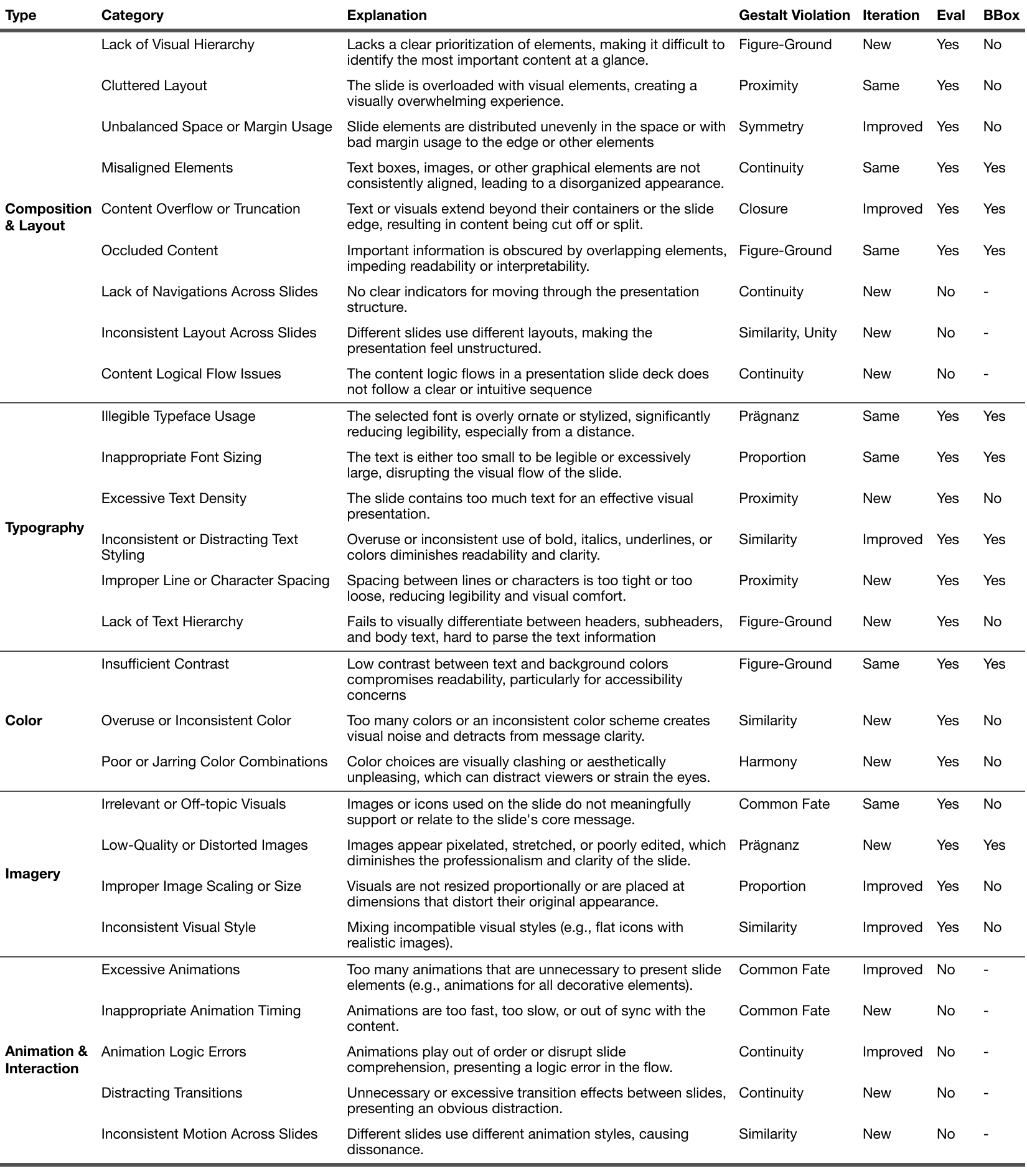}
    \caption{The \codename taxonomy of slide design flaws, organized by category, with explanations, Gestalt principle violations, iteration status (New: surfaced after design study; Improved: iterated our design study; Same: from original taxonomy version), evaluation inclusion (whether it is included in our evaluation study scope), and bounding box applicability.}
    \label{fig:taxonomy_table}
\end{figure*}

\subsection{Results}

To initiate the taxonomy development process, the research team conducted three internal workshops to brainstorm and document common slide design issues. While surface-level flaws such as misaligned text or low contrast are readily recognized, the team aimed to formalize these and less obvious problems into a structured taxonomy. This effort resulted in an initial draft that included 14 categories across three high-level groups: Layout and Organization, Text Formatting, and Multimedia. 

Over the course of the study, participant feedback led to significant improvements in both the content and framing of the taxonomy. First, category names and descriptions were refined for clarity, with several vague terms replaced by more precise language. For example, participants flagged ambiguous terms like ``Visual Overload or Underuse'' and recommended splitting or rephrasing them to improve interpretability and granularity.

Beyond categorical refinement, a more significant shift occurred in how flaws were conceptualized. Instead of solely capturing errors introduced by slide creators, the revised taxonomy integrates visual perception principles to account for \textit{how audiences interpret slide content}. In particular, our expert's feedback connected Gestalt principles \cite{wertheimer2017untersuchungen} to our taxonomy. For example, ``Cluttered Layout'' relates directly to the Proximity principle, as it disrupts the viewer's ability to perceive which elements belong together. ``Misaligned Elements'' violates the Continuity principle by breaking the natural flow that our eyes want to follow. ``Inconsistent or Distracting Text Styling'' contradicts the Similarity principle, where related text elements should share visual attributes to signal their relationships. ``Insufficient Contrast'' undermines Figure-Ground perception, making it difficult to differentiate foreground content from background elements. Using a linguistic metaphor~\cite{winograd1986understanding}, the initial taxonomy focused on \textit{lexical} and \textit{syntactic} correctness, while the revised version incorporates \textit{semantic} and \textit{pragmatic} dimensions---that is, how visual structure contributes to meaning and context.

The final taxonomy, presented in Figure~\ref{fig:taxonomy_table}, comprises 27 detailed flaw categories grouped under five high-level dimensions: Composition \& Layout, Typography, Color, Imagery, and Animation \& Interaction. Compared to the initial version, the refined taxonomy incorporates feedback that led to new categories (e.g., Illegible Typeface Usage, Lack of Visual Hierarchy), clearer boundaries between overlapping issues (e.g., Cluttered Layout vs. Excessive Text Density), and terminology grounded in visual design theory \cite{lidwell2010universal,ware2019information,nielsen1994enhancing}. Note that in our evaluation, we picked 19 categories instead of all 27 (``Eval'' column in Figure~\ref{fig:taxonomy_table}) because the unselected categories represent flaws across slides or dynamic ones like animation. They are, therefore, out of our current research scope.

\section{Compiling a Dataset of Slide Design Flaws}

To operationalize the developed taxonomy, we collected and annotated a dataset of presentation slides based explicitly on our design flaw categories. Although existing slide datasets offer useful resources for visual analyses, none currently provide structured annotations indicating specific visual design problems. Our annotated dataset therefore uniquely supports the structured evaluation of automated systems designed to detect concrete design flaws. The dataset is released and open-sourced for the community.\footnote{https://github.com/zhuohaouw/SlideAudit}

\subsection{Data Collection}
To construct the \codename dataset, we curated and augmented a corpus of 2400 presentation slides. These slides originate from three distinct sources, each selected to represent a different mode of slide creation and use: (1) a publicly available government slide deck dataset, (2) publicly shared slide decks from Google Slides, and (3) slides synthesized by Google Gemini using AI-generated prompts designed to elicit various layouts and content structures. From each source, we selected 200 slides, resulting in a total of 600 source slides.

To expand the dataset and create a controlled setting for evaluating design flaws, we automatically altered each original slide using three distinct alterations. These included changes to (1) within-object alignments, (2) between-object layouts, and (3) typography attributes such as font size and weights. These alterations were designed to introduce visual and structural flaws that commonly appear in real-world slides but may not be reliably present in unaltered slides. 

More specifically, the \textit{within-object alignment} alterations were achieved by targeting the internal structure of individual elements. These included changing text alignment (e.g., shifting from left-aligned to centered or right-aligned), and disrupting the internal coherence of elements such as charts, tables, and grids. We applied localized alterations such as rotations or slight displacements of subcomponents to intentionally break perceived alignment and internal order, thereby violating visual expectations of consistency and grid conformance. In contrast, \textit{between-object layout} alterations operated at the inter-object level, modifying the spatial relationships between entire elements. These included resizing objects to introduce disproportionate scaling, repositioning elements to disturb visual flow, and intentionally creating overlaps to simulate crowding. The goal was to disrupt the overall layout structure---breaking whitespace balance, violating margin conventions, and degrading the perceived hierarchy and compositional stability of the slide. Lastly, \textit{typography} alterations involved altering font properties such as size, weight, and spacing. These changes introduced readability and emphasis issues commonly observed in poor slide design, such as overly small or inconsistent text, or imbalanced visual weight across the slide. We include alteration examples in Appendix \ref{sec:alt_examples}.

The idea and choice of alterations were informed by preliminary reviews of prior literature \cite{wu2024uiclip} and by informal consultations with expert designers. Our goal was to simulate common but impactful design issues that may degrade readability, emphasis, or professional appearance. Importantly, alterations were applied to both visually optimal and already suboptimal slides, ensuring the dataset contains a range of quality levels. Each of the 600 original slides was altered using all three alteration strategies, resulting in 2400 distinct slides.

We avoided stochastic or arbitrary perturbations; instead, each alteration was implemented through predefined templates or rule-based scripts that ensured repeatability and internal consistency. For example, position changes involved shifting key objects, like slide titles, out of alignment, while layout changes disrupted common grid structures or slide symmetry. These systematic modifications made it possible to evaluate both human and automated identification of slide design flaws.

\begin{figure}
    \centering
    \includegraphics[width=\linewidth]{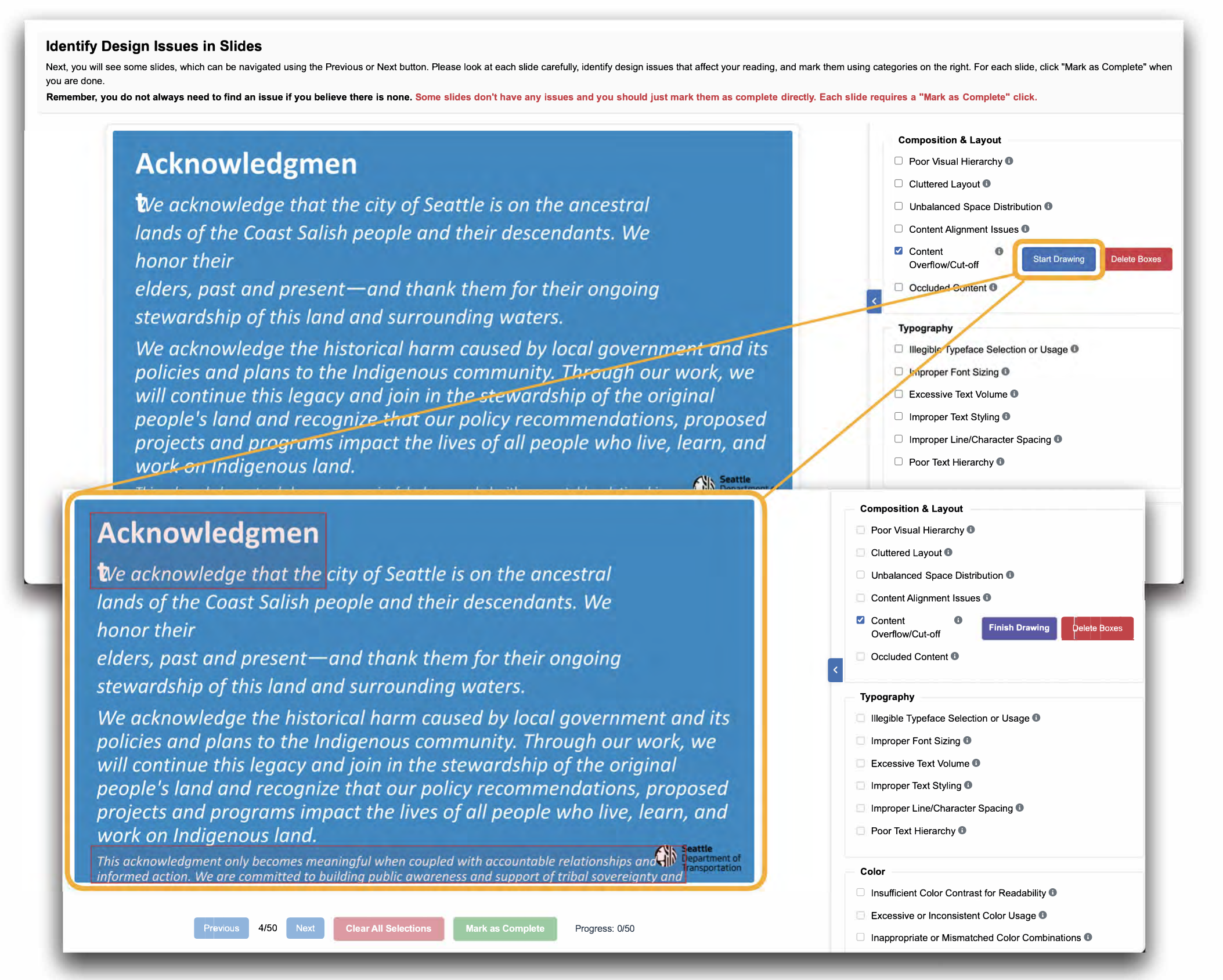}
    \caption{The \codename annotation interface. Annotators navigate slides, identify design flaws using the taxonomy categories on the right, and mark as complete when finished.}
    \label{fig:annotation_interface}
\end{figure}

\subsection{Dataset Annotation}
We annotated all 2400 slides (600 original slides and 1800 altered slides) in the \codename dataset using our design flaw taxonomy through a structured crowdsourcing process. This section outlines our annotation methodology, quality control procedures, and agreement analysis.

\subsubsection{Method}
\paragraph{Participants}
We recruited participants from \textit{Prolific}, a crowdsourcing platform that generally produces much higher quality results than other platforms like Amazon's Mechanical Turk. We required participants to (1) have prior experience with design or slide creation, and (2) hold at least a bachelor's degree. To ensure annotation quality, we embedded a qualification step---a tutorial, practice, and a quiz---before participants could access the main task. In total, 145 participants completed the annotation tasks and were approved for inclusion. An additional 15 participants completed the tasks but were rejected based on quality control, and 122 participants began but exited before completion of the tutorial section or passing the quiz. The approved participants completed the tasks for 78.7 minutes on average ($\sigma = 29.0$). For the participants who completed all tasks and gave their consent to report their demographics, 61.38\% have undergraduate degrees, 31.72\% have graduate degrees, and 6.9\% have doctoral degrees. Their ages ranged from 18 to 73 years ($\mu=33.6, \sigma=10.9$), with 44.1\% females and 55.9\% males. 

\paragraph{Apparatus}
We developed two custom tools to support the annotation process:

\begin{itemize}

\item \textit{Qualification Quiz Interface}: This tool served as both a tutorial and a screening mechanism. The survey via Qualtrics introduced our taxonomy in a scaffolded manner, presented examples of similar and potentially confusing design flaw types, and trained users on drawing bounding boxes to locate those flaws. Participants were required to pass a multiple-choice quiz identifying obvious design flaws in 10 sample slides. Only those who correctly answered all questions received a password granting access to the annotation task.

\item \textit{Custom Web Annotation Tool}: We developed a web-based annotation interface (Figure~\ref{fig:annotation_interface}) to collect structured labels and localized bounding boxes for flaw categories. Participants used the tool to: (1) select relevant flaw labels from the taxonomy, and (2) optionally draw bounding boxes to spatially locate flaws. Bounding boxes were enabled only for a predefined subset of nine taxonomy categories (see Column ``BBox'' in Figure~\ref{fig:taxonomy_table}), selected based on their clear spatial localization within the slide (e.g., misaligned elements, illegible typefaces). We intentionally omitted bounding box annotations for the remaining categories, as these typically involved holistic or relational issues (e.g., lack of visual hierarchy) that cannot be meaningfully confined to specific spatial regions. The annotation interface supported canvas-based interaction, zooming controls, and integrated definitions of each flaw category from our taxonomy.

\end{itemize}

\paragraph{Procedure}
Each approved participant was assigned 50 slides for annotation. For each slide, participants reviewed the full-slide image and were asked to select any applicable flaw categories; for certain categories (e.g., overlapping objects, low contrast text), they also drew bounding boxes indicating the affected regions.

Participants were instructed to annotate only what they could confidently identify. We designed the interface to prioritize accuracy over speed and provided instructions discouraging over-labeling. The bounding box option was only available for object-level categories and disabled for global or stylistic ones (e.g., poor visual hierarchy).

As a quality control step, we sampled 10 slides from each participant's submission and used an LLM-based pipeline to flag suspicious entries (e.g., no annotations applied, all categories selected indiscriminately, or incoherent bounding boxes). Six participants' submissions failing this check were manually reviewed and rejected. Finally, each slide in the dataset received annotations from three independent annotators.

\subsubsection{Analysis}
We aggregated the categorical annotations using a majority vote scheme: if at least two of three annotators selected a given flaw label for a slide, we marked that label as genuinely present. Labels selected by all three annotators were flagged as ``strong agreement'' cases.

To assess inter-rater reliability across the three annotators per slide, we computed Fleiss' $\kappa$, which resulted in a score of 0.26, indicating fair agreement. While this score reflects moderate variability, it aligns with prior work UICrit (Fleiss' $\kappa = 0.29$) in design evaluation where subjectivity and interpretation are expected. 

For bounding boxes, we collected annotations for nine categories where design flaws could be spatially identified. A bounding box was retained if at least two annotators provided overlapping regions for the same category. 

\begin{figure}
    \centering
    \includegraphics[width=0.8\linewidth]{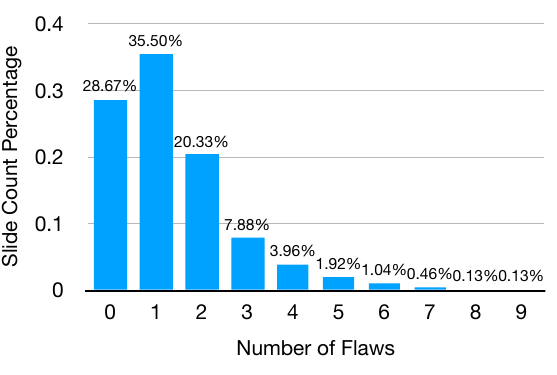}
    \caption{Distribution of the number of design flaws per slide in the \codename dataset (N=2400)}
    \label{fig:flaw_per_slide}
\end{figure}

\subsection{Dataset Summary}

We summarize the statistics of the \codename dataset across all 2,400 annotated slides. Among them, 28.7\% of slides were labeled with zero flaws, 35.5\% with one flaw, 28.2\% with two or three flaws, and 7.6\% with four or more flaws (Figure~\ref{fig:flaw_per_slide}). The most extreme cases included three slides with nine annotated flaws. On average, each slide contained 1.37 flaws ($SD=1.38$).

Across the four high-level categories, \textit{Composition \& Layout} was the most frequently occurring flaw type, appearing in 70.5\% of all slides, followed by \textit{Typography} (43.0\%), \textit{Color} (13.7\%), and \textit{Imagery} (9.6\%). The relatively low frequency of \textit{Imagery}-related flaws may be due to the increased cognitive demand required to interpret visual content and assess its relevance, leading annotators to focus more heavily on spatial and structural issues.

\begin{figure}
    \centering
    \includegraphics[width=\linewidth]{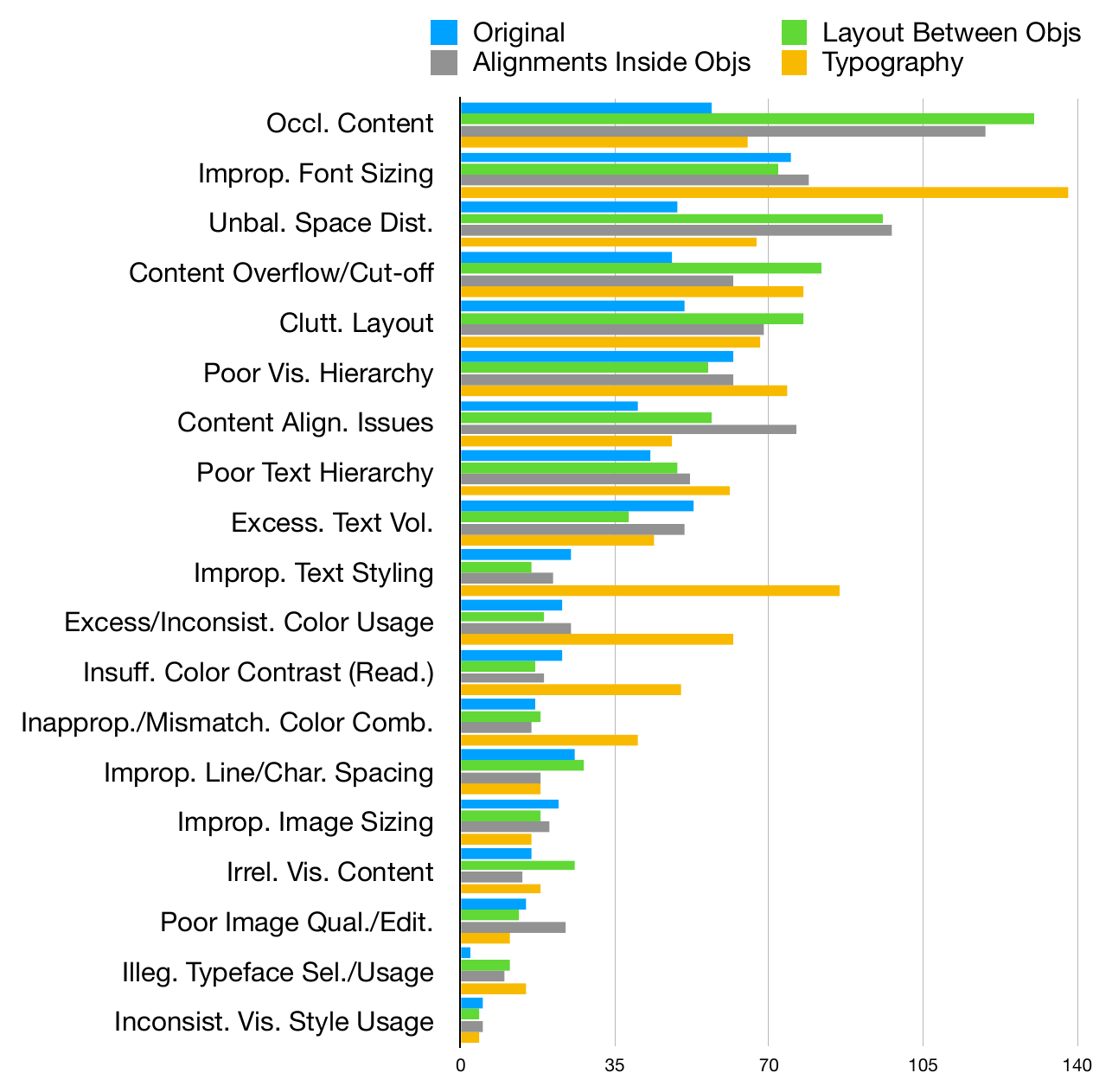}
    \caption{Frequency of slide design flaws across original slides and three alteration types in the \codename dataset. Each flaw is represented by four bars indicating its occurrence in the original slides and in three slide alterations. Each alteration group consists of 600 slides.}
    \label{fig:flaw_occurrences}
\end{figure}

At the category level, \textit{Occluded Content} (15.5\%) and \textit{Inappropriate Font Sizing} (15.2\%) were the most frequently identified flaws, while \textit{Inconsistent Visual Style} appeared least often, occurring in only 0.75\% of slides. To better understand how different types of slide alterations contributed to these design issues, we analyzed the distribution of flaw categories across alteration types (Figure~\ref{fig:flaw_occurrences}). As expected, typography-related alterations led to substantially higher frequencies of flaws such as improper font sizing, inconsistent text styling, poor color usage, low contrast, and mismatched color combinations. Similarly, layout and alignment alterations were strongly associated with increased occurrences of flaws like occluded content, unbalanced space or margin distribution, content overflow, and misaligned elements.

\section{Evaluation}
We conducted three different evaluation studies: (1) Performance evaluation on how LLMs identify design flaws, (2) an evaluation using an existing design critique framework, and (3) a remediation study to show whether our taxonomy helps LLMs fix problematic slides better.

\subsection{Performance Evaluation}
\label{sec:performance}
To assess how effectively current LLMs can identify design flaws in slides, we conducted a systematic evaluation across three model providers, four prompting strategies, and two input modalities, aiming to benchmark LLM performance under varied prompting conditions and input representations. 

\subsubsection{Models and Input Variants}
We selected three leading commercial LLMs representative of the current state of the market: GPT-4o (gpt-4o-2024-08-06) by OpenAI \cite{gpt4o}, Claude 3.7 Sonnet (claude-3-7-sonnet-latest) by Anthropic \cite{claude37sonnet}, and Gemini 2.0 Flash by Google \cite{gemini}. Each model was evaluated with two types of input: (1) Raw slide-as-image only, and (2) slide-as-image with a structured object description. This representation included a list of visible objects with their types (e.g., text box, image), spatial coordinates, and object-level IDs, supplementing visual information. 

\subsubsection{Prompting Strategies}
We developed four prompting strategies to test the models under increasingly informative conditions:

\begin{enumerate}
    \item 
    \textit{Baseline Prompting (Zero-Shot)}: This variant involved a direct query to the model using only the image (and optionally, object structure) with minimal instruction (Appendix~\ref{sec:prompts} in supplemental materials). This prompt did not contain any reference to the taxonomy as a control condition to evaluate how well models could surface slide design issues without any guiding knowledge.
    
    \item 
    \textit{High-Level Category Prompting}: This variant augmented the baseline with four top-level taxonomy categories. These were presented as areas of focus, prompting the model to consider each when evaluating the slide. This approach tested whether lightweight categorical scaffolding could help models surface a more structured range of issues.
    
    \item 
    \textit{Full Taxonomy Prompting}: We then injected our complete taxonomy into the prompt, including all 19 categories with their corresponding definitions. The prompt also instructed the model to identify all applicable categories from the taxonomy and provide a brief explanation for each. This tested the model's ability to map perceived flaws directly to taxonomy definitions.
    
    \item 
    \textit{Taxonomy + Computational Augmentation}: The fourth variant included the full taxonomy along with three additional computational signals designed to approximate how visual saliency, contrast, and perceptual grouping might inform human judgments of slide quality:
    
    \begin{itemize}
        \item \textit{Visual Saliency Map}: Generated using the Unified Model of Saliency and Importance (UMSI++) \cite{fosco2020predicting, jiang2023ueyes}, this heatmap indicated the predicted visual attention distribution on the slide. It provided the model with a potential approximation of visual hierarchy.
        \item \textit{Computational Grouping}: We applied a computational model of Gestalt principles based on persistent homology \cite{chen2024gestalt}, which outputs a segmentation of perceptual groups. This context was included to help the model consider whether slide elements adhered to or violated perceptual organization norms such as proximity and alignment.
        \item \textit{Color Metrics}: We calculated a slide-level colorfulness score \cite{hasler2003measuring} and, for each text-containing object, the WCAG 2.1 contrast ratio \cite{w3c_wcag2.1} between foreground and background colors. These metrics, similar to those used in accessibility checkers \cite{AndroidAccessibilityTesting}, helped ground model judgments of readability and visual noise.
    \end{itemize}

\end{enumerate}

\begin{table*}[h]
  \centering
    \caption{F1 scores of three models across four augmentation strategies and two input types. GPT-4o shows the strongest performance overall.}
  \label{tab:model_performance}
  \begin{tabular}{@{}l@{\hspace{8pt}}cc@{\hspace{8pt}}cc@{\hspace{8pt}}cc@{\hspace{8pt}}cc@{}}
    \toprule
    \multirow{2}{*}{\textbf{Model}} 
    & \multicolumn{2}{c}{\textbf{Baseline}} 
    & \multicolumn{2}{c}{\textbf{Category}} 
    & \multicolumn{2}{c}{\textbf{Taxonomy}} 
    & \multicolumn{2}{c}{\textbf{Tax. + Comp. Augment.}} \\
    
    & Raw Image & W/ Obj. Desc. 
    & Raw Image & W/ Obj. Desc. 
    & Raw Image & W/ Obj. Desc. 
    & Raw Image & W/ Obj. Desc. \\
    \midrule
    \textbf{claude-3-7} & 0.492 & 0.519 & 0.518 & 0.514 & 0.589 & 0.612 & 0.568 & 0.583 \\
    \textbf{gemini-2.0-flash} & 0.515 & \textbf{0.531} & 0.543 & 0.554 & 0.587 & 0.604 & 0.568 & 0.597 \\
    \textbf{gpt-4o} & 0.476 & 0.476 & \textbf{0.586} & 0.584 & \textbf{0.634} & \textbf{0.655} & \textbf{0.616} & \textbf{0.631} \\
    \bottomrule
  \end{tabular}
\end{table*}

\begin{table*}[h]
  \centering
  \caption{Precision, Recall, and F1 scores comparing baseline and taxonomy-based approaches.}
  \label{tab:precision_recall_f1}
  \begin{tabular}{@{}l@{\hspace{10pt}}ccc@{\hspace{10pt}}ccc@{\hspace{10pt}}ccc@{}}
    \toprule
    \multirow{2}{*}{\textbf{Model}} 
    & \multicolumn{3}{c}{\textbf{Precision}} 
    & \multicolumn{3}{c}{\textbf{Recall}} 
    & \multicolumn{3}{c}{\textbf{F1}} \\
    
    & Baseline & Taxonomy & $\Delta$ 
    & Baseline & Taxonomy & $\Delta$ 
    & Baseline & Taxonomy & $\Delta$ \\
    \midrule
    \textbf{Claude} & 0.533 & 0.629 & +0.096 & 0.575 & 0.600 & +0.025 & 0.519 & 0.612 & +0.093 \\
    \textbf{Gemini} & 0.540 & 0.646 & +0.106 & 0.528 & 0.586 & +0.058 & 0.531 & 0.604 & +0.073 \\
    \textbf{GPT}    & 0.455 & 0.668 & +0.213 & 0.500 & 0.645 & +0.145 & 0.476 & 0.655 & +0.179 \\
    \bottomrule
  \end{tabular}
\end{table*}

\subsubsection{Evaluation Method} We discuss how we selected a subset of slides for evaluation, how we mapped free-text outputs to our taxonomy in baseline and category prompting, and how we tested the reliability of consistency across different LLM runs.

\textbf{Selecting an Evaluation Set}: For each slide, we queried each model across all prompting variants and both input types, yielding 24 conditions per slide. To manage the evaluation scope while maintaining representative coverage, we applied stratified random sampling to the full 2400 slides. Slides were grouped into three strata based on the number of annotated flaws: slides with (1) 0–1 issues, (2) 2–3 issues, and (3) 4 or more issues. We randomly sampled 50 slides from the first group (``good'' slides), 100 from the second (``moderate''), and 50 from the third (``highly flawed''). This resulted in a final evaluation set of 200 slides, each tested under all 24 \textit{model–prompt–input} combinations. Besides asking LLMs to identify flaws, we explicitly instructed the models not to over-interpret or fabricate issues if a slide appeared well-structured, to mitigate over-detection.

\textbf{Mapping Free-Text Response to Taxonomy}: For the first two prompting variants, where the full taxonomy was not provided, we obtained model outputs as free-text responses. To standardize comparisons across variants, we employed a separate ``result comparator'' LLM to categorize these free-text-identified flaws according to the defined taxonomy. To verify the reliability of this automated comparator approach, we manually coded a randomly selected subset of 36 free-text responses (1\% of all free-text outputs), independently mapping each response onto the taxonomy categories. This reliability check yielded substantial human–LLM agreement (Cohen's $\kappa = 0.68$). Upon further qualitative analysis, we observed that disagreements arose predominantly from ambiguously described design flaws in free-text responses (e.g., stating ``an image placement is awkward'' without specifying the underlying flaw) or out-of-scope issues unrelated to design flaws (e.g., grammar errors). These types of disagreements minimally impact the accurate identification and categorization of clearly articulated design issues. Therefore, we believe that the task of taxonomy-mapping from free-text responses is well-defined and that the comparator LLM achieves robust performance.
However, given the subjective nature of this approach, larger-scale validation through human annotators may be required in future work.

\textbf{Reliability Test}: Given the nature of the subjective design evaluation task, we also tested the consistency of different LLM runs. After we got the 24 conditions' results, we ran the best-performing LLM (GPT-4o) three times on the same set. We performed an analysis of variance based on mixed logistic regression \cite{gilmour1985analysis, stiratelli1984random} on the three runs and found no detectable difference in predictions, $\chi^2(2, N\mathord{=}11,400)=1.46$, $p=.48$. Given the large amount of data analyzed, any meaningful difference in output would likely have been detected; therefore, we found the model output to be consistent
across the various runs.

After we did a full performance evaluation on the test set of 200 slides, we picked the best-performing LLM variant (model, input configuration, and prompting technique) to evaluate the whole set of 2400 slides. 

\textbf{Ad-hoc Analysis of Taxonomy Coverage}: To further assess the coverage of our taxonomy, we conducted an ad-hoc analysis of the raw outputs generated by the LLMs across all prompting conditions. We manually inspected cases where the model flagged potential issues but no existing taxonomy label was assigned. This review did not reveal any novel issue types outside our defined categories, suggesting that the current taxonomy sufficiently captures the range of slide design flaws surfaced in our experiments.

\subsubsection{Evaluation Metrics}
We report macro-averaged precision, recall, and F1 scores across the 19 flaw categories, with a focus on F1 score. We treat each category equally. In the meantime, most slides contain only a small subset of possible flaws, resulting in a highly imbalanced label distribution where the majority of category labels are negative (i.e., no flaw present). In this context, F1 score serves as a balanced metric, capturing both the model's ability to detect real issues (recall) and to avoid over-identifying problems that are not present (precision). We compute the macro F1 score of $C=19$ categories using the arithmetic mean value of F1 scores per category:

\begin{equation}
    \text{Macro F1} = \frac{1}{C} \sum_{i=1}^{C} F1_i
\end{equation}

\subsubsection{Results}

\textbf{Evaluation Set}: Across all 24 model–prompt–input configurations, macro F1 scores ranged from 0.476 to 0.655 (Table~\ref{tab:model_performance}). The best-performing variant was GPT-4o using image input with structured object description and full taxonomy prompting, achieving a macro F1 score of 0.655.

Performance consistently improved as the level of prompt guidance increased. From the baseline (zero-shot) condition to high-level category prompting and then to full taxonomy prompting, all three LLMs showed gains in both precision and recall. Table~\ref{tab:precision_recall_f1} reports the deltas between baseline and taxonomy conditions, with GPT-4o showing the largest F1 improvement of +0.179, followed by Claude (+0.093) and Gemini (+0.073). This trend highlights the value of structured, domain-specific prompting when directing LLMs to perform slide critique tasks.

On the other hand, the fourth prompting strategy (full taxonomy with computational augmentation) did not lead to further improvements. In some cases, performance slightly decreased compared to full taxonomy alone. A possible explanation is that providing additional saliency, color, and Gestalt-based signals may have inadvertently caused LLMs to over-focus on those specific aspects, leading to false positives. This aligns with known tendencies of LLMs to over-interpret provided context, suggesting the need for more controlled or targeted augmentation strategies when combining symbolic and perceptual inputs.

Additionally, evaluating the comparator LLM indicated a propensity to frequently include ambiguous or out-of-scope flaws. Although the comparator LLM was implemented specifically to map such responses into our defined categories, effectively standardizing and enhancing baseline and high-level prompting outputs, performance of these variants still fell behind the explicit full-taxonomy prompting approach. This reinforces the value of maintaining an explicit, clearly defined taxonomy: even with \textit{post hoc} enhancement to categorize ambiguous responses, initial precision in prompting directly with the comprehensive taxonomy remains advantageous.

\textbf{Full Dataset}: When applying the best-performing configuration (GPT-4o, image + object description, full taxonomy) to the \textit{full dataset} of 2400 slides, the resulting category-level scores were: Macro-averaged Precision = 0.579, Recall = 0.577, and F1 = 0.578. The slight decrease compared to the evaluation subset may be attributed to differences in slide distribution, as the test set was stratified to ensure representation across a range of flaw densities.

Overall, the results demonstrate that while prompting strategies and structured input improve model behavior, the task of identifying slide design flaws remains highly interpretive. LLMs can be made to exhibit reasonable consistency but still fall short of matching human-annotated results. 

\begin{figure*}[t]
    \centering
    \includegraphics[width=0.85\textwidth]{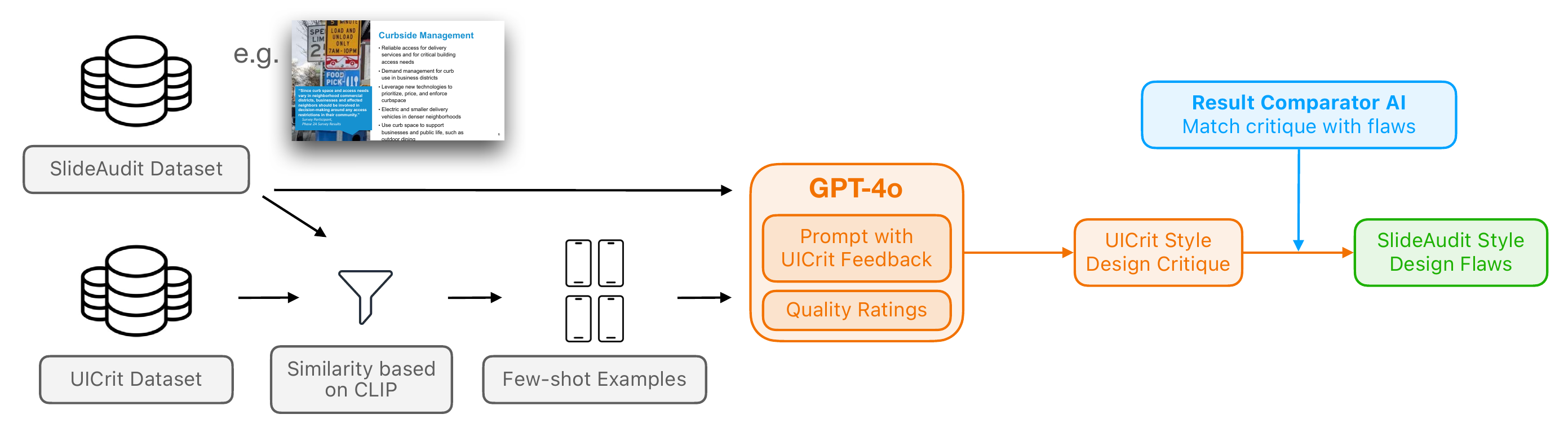}
    \caption{Using \codename data in UICrit pipeline. Few-shot examples were generated using CLIP similarity and GPT-4o was prompted with design critique examples to provide comments on problematic slides. The result is evaluated using the same result comparator AI.}
    \label{fig:uicrit_pipeline}
\end{figure*}

\textbf{Bounding Boxes}: Besides identifying flaws in slides, it is also important to ground these flaws in the image, if possible. However, grounding bounding boxes in such design evaluation tasks is extremely challenging \cite{duanUICritEnhancingAutomated2024}. We further evaluated LLM's ability to localize those issues by generating bounding boxes. Specifically, we assessed bounding box quality by computing the intersection-over-union (IoU) between the predicted box and the ground truth box for each correctly identified flaw (true positives only). False positive detections were excluded from this analysis, as no ground truth bounding boxes existed for comparison.

Using the best-performing configuration (GPT-4o with image + object description + full taxonomy), we observed an average IoU of 0.248 across all valid predictions. In prior related work, a similar task reported a best IoU score of 0.222 \cite{duanUICritEnhancingAutomated2024} among different prompting methods, which is close to what we evaluated. These data highlighted that generating accurate bounding boxes remains a notable challenge. Despite improvements in identifying category-level flaws, grounding those issues spatially on slides requires finer-grained visual reasoning that LLMs currently struggle to perform. This limitation suggests a need for better multimodal grounding strategies or complementary vision modules to support explainable design evaluation tasks.

\subsection{Evaluating with Prior Design Critique Frameworks}

Prior work has provided comparable materials in design critiques. CrowdCrit \cite{luther2015structuring} proposed seven design principles as a high-level taxonomy for design critiques of graphics. UICrit \cite{duanUICritEnhancingAutomated2024} described an LLM-based pipeline for automated UI evaluation. It generated natural language critiques for mobile user interfaces using few-shot learning with curated examples from an expert-annotated dataset. To compare \codename with these prior design critique works, we conducted two additional evaluations: (1) We evaluated two additional LLM variants similar to our evaluations in section \ref{sec:performance} to compare \codename's and CrowdCrit's taxonomies; (2) we adapted UICrit's automated evaluation pipeline to our setting by using \codename's dataset.

\subsubsection{Method} We introduce the two comparisons with prior work separately below:

\textbf{Comparing with CrowdCrit's Taxonomy}:
We incorporated CrowdCrit's seven high-level principles~\cite{luther2015structuring} and a set of 70 design critique statements into our performance evaluation pipeline. We constructed two additional prompting variants for this purpose. In the first variant, we augmented the baseline prompt in section \ref{sec:performance} with the seven CrowdCrit principles---\textit{Layout}, \textit{Readability}, \textit{Simplicity}, \textit{Emphasis}, \textit{Balance}, \textit{Consistency}, and \textit{Appropriateness}---each paired with a brief description to guide the model's interpretation. In the second variant, we included 70 critique statements that we generated to reflect the types of feedback represented in CrowdCrit, since the original paper did not release the full set of human-authored critiques. In this comparison, we used the best-performing LLM variant in the above experiment (GPT-4o with object descriptions). These generated critiques were formatted as natural language input to help the model identify relevant design issues. Both prompting conditions followed the same structure and evaluation procedures as our \codename category- and taxonomy-based methods, enabling consistent comparisons across approaches. 

\textbf{Evaluating with UICrit's Pipeline}:
In our adaptation, we treated slides as a form of static visual interface and applied the UICrit pipeline (Figure~\ref{fig:uicrit_pipeline}) directly to our slide dataset. We took the same 200 subset and used OpenAI's CLIP model \cite{radford2021learningtransferablevisualmodels} to compute visual similarity scores between each target slide and the mobile UI screenshots in the UICrit and RICO datasets \cite{deka_rico_uist17}. For each slide, we selected the eight most visually similar UI examples. These critiques were concatenated to form a prompt, which was then similarly provided to the best-performing LLM (GPT-4o with object descriptions) to generate a design critique for the target slide. This process mirrors UICrit's few-shot and visual-prompting approach. Unlike our taxonomy-guided evaluations, this comparison was limited to textual critique generation. We did not include bounding box evaluations.

To enable standardized evaluation, we applied the same result comparator AI used in our aforementioned study to match the generated critiques with categories in our slide flaw taxonomy. Doing so allowed us to measure how well the adapted UICrit pipeline aligned with our annotation framework and to evaluate whether knowledge learned from UI critique tasks could generalize to the domain of slide design evaluation.

\subsubsection{Results} We report on the evaluation results of each comparison below (Table \ref{tab:prior-framework-evals}):

\begin{table}[t]
\centering
\caption{Macro F1 Scores for Evaluation with Prior Design Critique Frameworks}
\label{tab:prior-framework-evals}
\begin{tabular}{@{}l c@{}}
\toprule
\textbf{Evaluation Variant} & \textbf{Macro F1} \\
\midrule
CrowdCrit (7 Principles) & 0.577 \\
CrowdCrit (Generated 70 Critiques) & 0.331 \\
UICrit Pipeline & 0.477 \\
\bottomrule
\end{tabular}
\end{table}

\textbf{CrowdCrit Taxonomy Evaluation}:
When prompted with the seven high-level principles, the best-performing LLM achieved a macro F1 score of 0.577---slightly below the performance of our taxonomy-guided prompting (F1 = 0.586). When prompted with the 70 generated critique statements simulating CrowdCrit's feedback style, the performance dropped substantially, with a macro F1 score of 0.331 (Table \ref{tab:prior-framework-evals}).

This result suggests that while CrowdCrit's high-level principles can guide LLMs to surface relevant slide issues, their broad and human-centered framing may be less effective than our targeted, category-driven taxonomy for automated evaluation. The lower performance of the critique-statement variant likely stems from the increased prompt length and semantic variability, which may have led to over-generation of issues and higher false positive rates. Together, these findings underscore the value of \codename's compact, machine-actionable taxonomy for benchmarking automated slide critique systems.

\textbf{UICrit Pipeline Evaluation}:
The pipeline adapting the UICrit dataset and framework achieved a macro F1 score of 0.477 (Table \ref{tab:prior-framework-evals}). Compared to the performance benchmarks in our earlier evaluation (Section~\ref{sec:performance}), these F1 score result is comparable to the baseline (zero-shot) prompting levels, but fall notably short of the best-performing taxonomy-guided configurations.

This outcome suggests that while LLM-based design critique methods developed for user interface evaluation can generate plausible feedback on slides, their ability to identify slide-specific flaws remains limited when applied out-of-domain. The UICrit framework was trained and tuned for mobile UI layouts and heuristics, which may not directly translate to the spatial and communicative conventions of presentation slides.

While this experiment offers preliminary evidence of low cross-media transferability, a deeper analysis of error types, critique phrasing, and semantic mismatches is needed to fully understand the gap. We encourage future investigations in this domain.

\subsection{Slide Remediation Study}

Now that we have evaluated whether LLMs can identify design flaws from presentation slides, we want to further investigate how effectively LLMs can generate actionable slide improvement strategies. Particularly, we want to evaluate whether our taxonomy meaningfully enhances this ability. We conducted a slide remediation study to answer this.

\subsubsection{Method}

We randomly sampled 50 slides from the previously described 200-slide evaluation set. For each slide, we selected two sets of flaw outputs generated by GPT-4o: (1) baseline zero-shot prompting and (2) full taxonomy-informed prompting. These two conditions represent opposite ends of the diagnostic quality spectrum.

Each identified flaw set was subsequently provided as input to GPT-4o to produce a step-by-step improvement plan (three examples are provided in Appendix \ref{sec:fix_example}). To validate the appropriateness and coherence of the GPT-generated proposals, we manually assessed a randomly selected subset (20 plans from 10 slides, covering both prompting conditions). All manually inspected plans were deemed clearly understandable and directly aligned with the corresponding diagnostic inputs.

We then recruited three participants who did not have prior knowledge in the taxonomy that we developed. They were only told to evaluate two comparable ``fix plans'' on a series of problematic slides with given guidelines. All three participants were Master's degree students from HCI and Design programs in our institution with extensive slide design experiences. 
To mitigate possible ordering bias in participant evaluations, we used a counterbalanced within-subjects design. Both improvement plans for each slide were presented side by side in random order, accompanied by the original slide. Three participants independently reviewed all 100 improvement plan variants (50 slides $\times$ 2 prompting conditions). Evaluation tasks required approximately two hours per participant.

Participants assessed each plan step on two dimensions:

\begin{itemize}
\item \textbf{Flaw Accuracy}: Validity of the identified flaw motivating this step.
\begin{itemize}
\item Accurate Flaw
\item Partially Accurate Flaw
\item Incorrect or Over-detection (False Positive)
\item Non-critical Flaw
\item Undetected Flaw (False Negative)
\end{itemize}
\item \textbf{Plan Executability}: Appropriateness and practicality of the suggested solution.
\begin{itemize}
\item Implementable Solution
\item Partially Implementable Solution
\item Generic or Ineffective Solution
\item Unnecessary Solution
\item Undetected Absent Solution
\end{itemize}
\end{itemize}

Participants also provided two holistic evaluations per slide:

\begin{itemize}
\item \textbf{Improvement}: Whether each individual plan significantly improved the original slide (\textit{binary rating: true or false for each of the two plans}).
\item \textbf{Preference}: Overall preferred plan, explicitly conditioned on improvement ratings:
\begin{itemize}
\item \textit{N/A}: Neither plan improved the slide (improvement = false for both).
\item Plan 1 or Plan 2: Exactly one plan improved the slide; automatically preferred.
\item Plan 1, Plan 2, or Equal: Both plans improved the slide, preference explicitly rated by the participant.
\end{itemize}
\end{itemize}

Responses were aggregated using majority voting across participants to determine the final step-level labels and holistic judgments. Agreement among raters at step-level analysis indicated moderate consistency, with Fleiss' $\kappa = 0.57$ for flaw accuracy and $\kappa = 0.52$ for plan executability, with an overall combined step-level agreement of $\kappa=0.56$. For holistic preference ratings, participant agreement was substantial, with a Fleiss' Kappa of $\kappa = 0.61$.

\begin{figure}
    \centering
    \includegraphics[width=\linewidth]{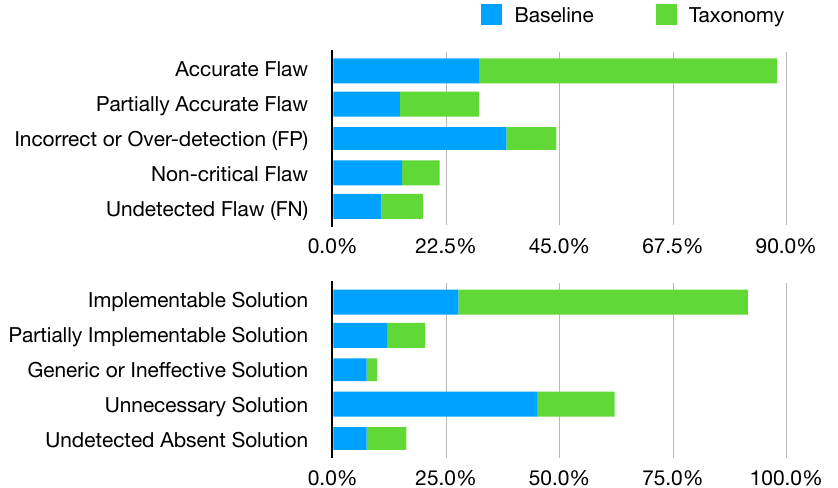}
    \caption{Comparison of fix plan quality between baseline and taxonomy-guided approaches. Taxonomy guidance yields more accurate flaw identifications and implementable solutions, with fewer false positives and unnecessary fixes.}
    \label{fig:fix_plan}
\end{figure}

\subsubsection{Results}

In total, 82.0\% (41 out of 50) of slides were judged as significantly improved by at least one of the two generated improvement plans. Note that even when slides' design flaws were not fully identified by LLMs, the identified parts and generated plans could still give participants a positive feeling that slides can be improved with the plans. Among the improved slides, participants preferred the taxonomy-informed plan in 87.8\% (36 out of 41) of cases, while 9.8\% (4 out of 41) exhibited no preference between the two approaches. Additionally, a special case occurred for one slide initially containing no substantial flaws, where the taxonomy-informed plan correctly produced no remediation steps and participants explicitly preferred this accurate response. These findings support the practical utility of taxonomy-informed prompting in guiding effective and contextually appropriate slide-improvement strategies.

As shown in Figure~\ref{fig:fix_plan}, the taxonomy-informed prompting improved both flaw identification accuracy and the executability of suggested solutions. Under the taxonomy-informed condition, the proportion of accurately identified flaws increased from 28.9\% (baseline) to 59.0\%. Conversely, false-positive identification (incorrect or over-detection) dropped markedly from 34.6\% to 9.6\%, and non-critical flaw detections decreased from 13.8\% to 7.2\%. Together, these changes demonstrate improvements in diagnostic precision.

For proposed fixes, the percentage of implementable solutions increased substantially from 27.7\% (baseline) to 63.9\%. Critically, taxonomy-informed prompting reduced unnecessary solutions from 45.3\% to 16.9\%, and generic or ineffective solution proposals decreased markedly from 7.5\% to 2.4\%. As unnecessary or generic recommendations can significantly degrade user experiences, these reductions reflect meaningful improvements toward context-specific and practical remediation guidance.

The improved diagnostic precision and higher executability of solutions have important implications in real-world assistive technology contexts, where presenting unnecessary or inaccurate suggestions may overload users and diminish trust. Taxonomy-informed plans proved more concise, actionable, and closely aligned with actual slide flaws, highlighting the practical value of integrating domain-specific guidance into LLM prompting strategies.

\section{Discussion}

We discuss the broader implications, dataset utility, limitations, and future work of \codename.

\subsection{Taxonomy Implications and Dataset Utility}

\codename is grounded in a rigorously developed taxonomy that reframes design flaws not simply as visual errors, but as perceptual and communicative breakdowns, shaping what viewers notice, how they interpret information, and how effectively a slide conveys its message. We see this taxonomy as a foundation for assistive systems that support, rather than override, human creativity. This approach may be particularly valuable for blind and visually impaired users, who have the same creative goals and could benefit from automated design evaluation and improvement when systems are designed with their needs in mind. Our aim is not to use AI to enforce rigid aesthetic standards, but to provide meaningful, context-sensitive feedback that helps users express their ideas more clearly and effectively.

In parallel, \codename offers a unique foundation for computational understanding of slide design. By combining a principled taxonomy, synthetically altered data, and rigorously executed annotations, it enables both benchmarking and training of models for design critique. Our evaluations highlight its value in testing LLM capabilities, adapting related frameworks, and exploring AI-assisted remediation. As the first dataset of its kind focused on visual clarity and communication effectiveness, \codename supports broader research in accessibility, slide automation, and AI-driven visual feedback. It also has a potentially broader application to sighted slide creators as well who have the needs to examine slide aesthetics before presentation.

\subsection{Limitations and Future Work}
Despite its contributions, our work has limitations. First, slide design evaluation remains inherently subjective, as design judgments often depend on context, intended audiences, and specific presentation goals. Although our taxonomy demonstrated effectiveness, evaluations indicated the need for novel LLM interaction paradigms designed specifically to subjective tasks like design and aesthetic assessments. For example, future LLM interactions could more effectively align with individual users' preferences by learning from limited examples or visual queries provided by the user. Consequently, our taxonomy offers a valuable foundation for grounding user preferences, enabling the future development of customized LLMs that provide personalized design evaluation and guidance. Given the same reason for the annotation task's subjectivity, annotating fine-grained design flaws required carefully trained annotators, making large-scale annotation resource-intensive. As discussed earlier, our dataset was built with strict quality control to ensure consistency, but future scaling may require more efficient semi-automated workflows. Additionally, while we drew slides from diverse sources---including real-world presentations and AI-generated content---potential biases in style, structure, or quality may still exist. These limitations offer opportunities for expanding source diversity and refining annotation protocols in future work.

Second, our evaluations did not isolate how specific computational augmentations affect detection performance across categories. Also, our remediation study was small in scale and relied upon participant judgment rather than real plan execution. Additionally, our taxonomy and annotations were limited to flaws observable within individual static slides. We did not address interactive elements, slide-to-slide transitions, or consistency across a full deck---dimensions that are crucial in real-world authoring but outside the scope of this work. 

Third, our evaluation involving UICrit was preliminary; we employed its evaluation pipeline primarily for comparative accuracy assessments without a deeper, qualitative exploration. Future work can be conducted to further compare UI and slide design issue detection. For example, we can apply the same few-shot retrieval methods in our own \codename dataset in contrast to using UICrit's dataset to see how the results would change. We can also compare SlideAudit with DesignChecker \cite{huh2024designchecker} on its usefulness of locating and fixing web design issues.

Other future directions include training specialized models on \codename to improve performance over current LLM baselines, which remain limited ($F1<0.7$). Classification of design issues is the first and critical step for automated evaluation. And beyond classification, models could also be developed for localized feedback and automatic remediation planning. Most importantly, we aim to integrate this work and use other open-ended critique frameworks into assistive authoring systems for blind and visually impaired users who need to create presentation-ready slides independently. Such tools, when integrated with existing assistive technologies \cite{zhang2023a11yboard, zhang2023developing, huh2024designchecker, huh2023genassist, zhang2023understanding, zhang2024charta11y, chang2024editscribe}, could help users independently identify and resolve visual design issues, enabling the creation of slides that are not only functionally complete but also visually effective and professionally presentable.

\section{Conclusion}
    As generative AI continues to influence digital content creation, the need for robust, automated tools that support high-quality visual design has become increasingly important. Effective visual communication through presentation slides remains a nuanced challenge, requiring attention to both aesthetic and functional aspects of design. In this work, we introduced \codename, the first structured dataset and taxonomy explicitly created to support automated evaluation of design flaws in presentation slides. By systematically modeling common design issues and establishing a principled annotation and benchmarking pipeline, this work lays the foundation for future research in AI-driven design critique and remediation.

\begin{acks}
This work was supported by an Apple Scholars in AI/ML
Ph.D. fellowship and the University of Washington Center for Research and Education on Accessible Technology and Experiences (CREATE). The contents of this work were developed under a grant from the National Institute on Disability, Independent Living, and Rehabilitation Research (NIDILRR grant \#90REGE0026-01-00). NIDILRR is a center within the Administration for Community Living (ACL), Department of Health and Human Services (HHS). The contents of work do not necessarily represent the policy of NIDILRR, ACL, or HHS, and one should not assume any endorsement by the United States federal government.
\end{acks}

\bibliographystyle{ACM-Reference-Format}
\bibliography{ref}

\clearpage
\appendix

\section{Alteration Examples}
\label{sec:alt_examples}
\begin{figure*}[t]
    \centering
    \includegraphics[width=0.7\textwidth]{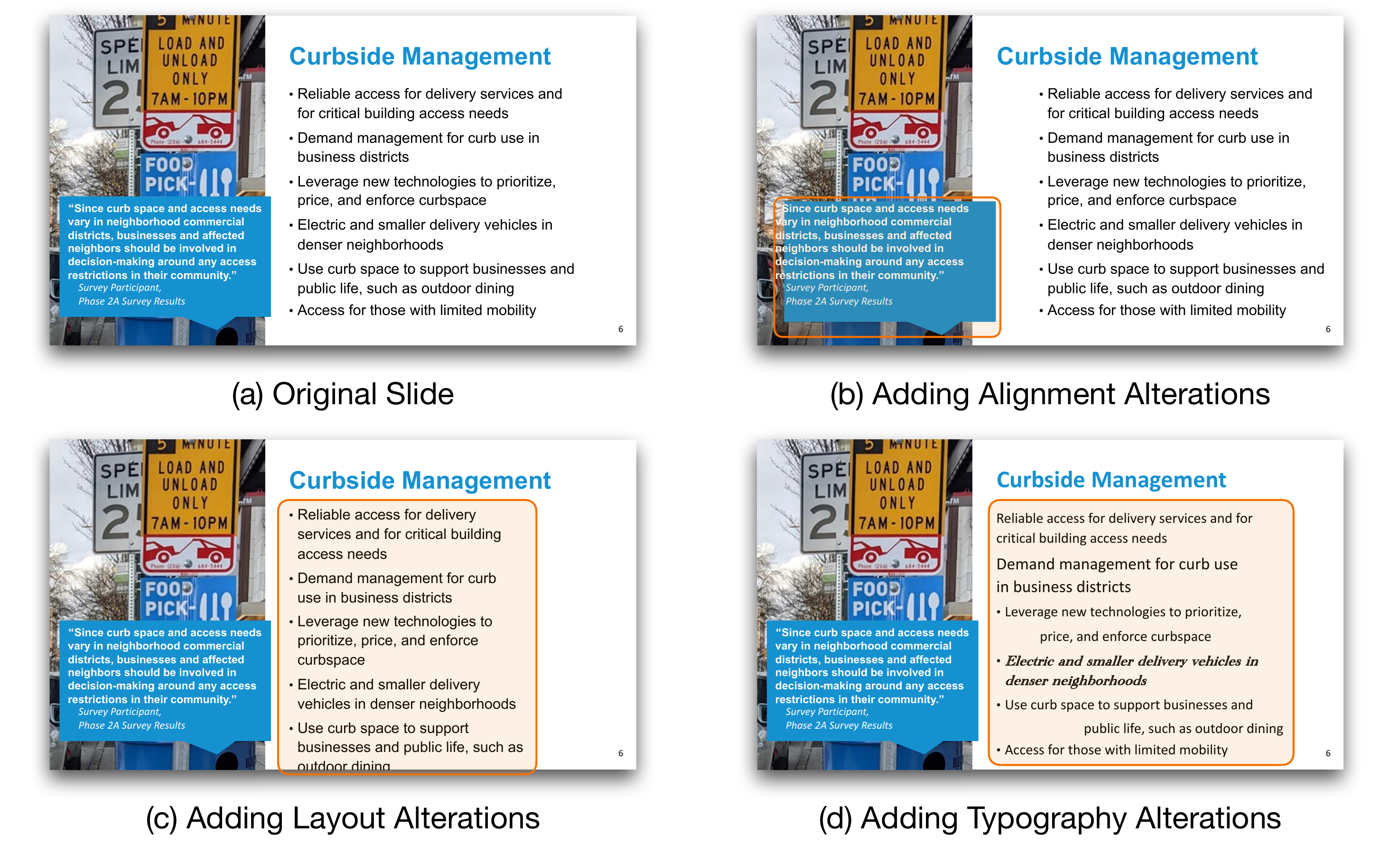}
    \caption{\codename dataset alteration example 1.}
    \label{fig:alt_eg_1}
\end{figure*}

\begin{figure*}[t]
    \centering
    \includegraphics[width=0.7\textwidth]{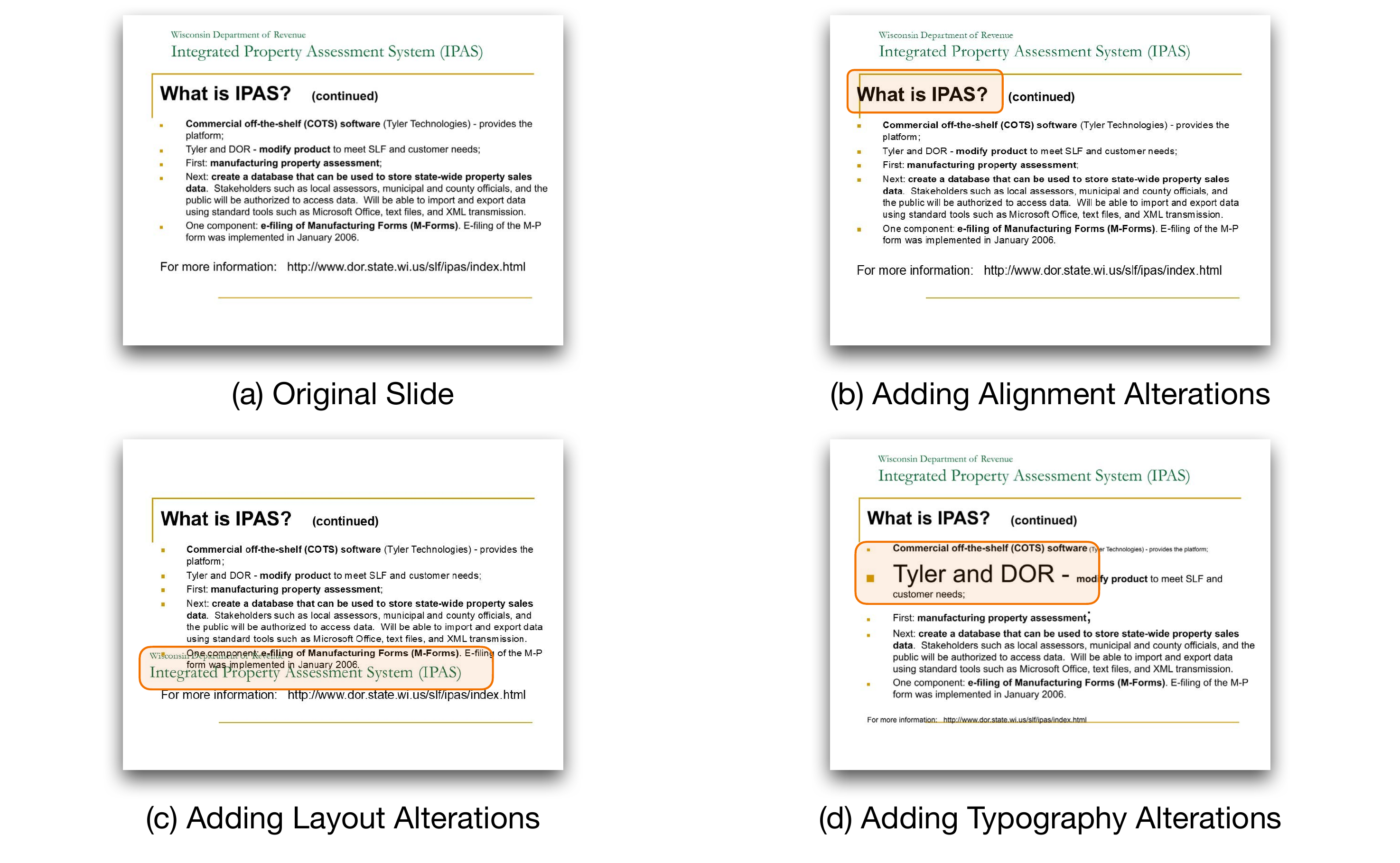}
    \caption{\codename dataset alteration example 2.}
    \label{fig:alt_eg_2}
\end{figure*}

\begin{figure*}[t]
    \centering
    \includegraphics[width=0.7\textwidth]{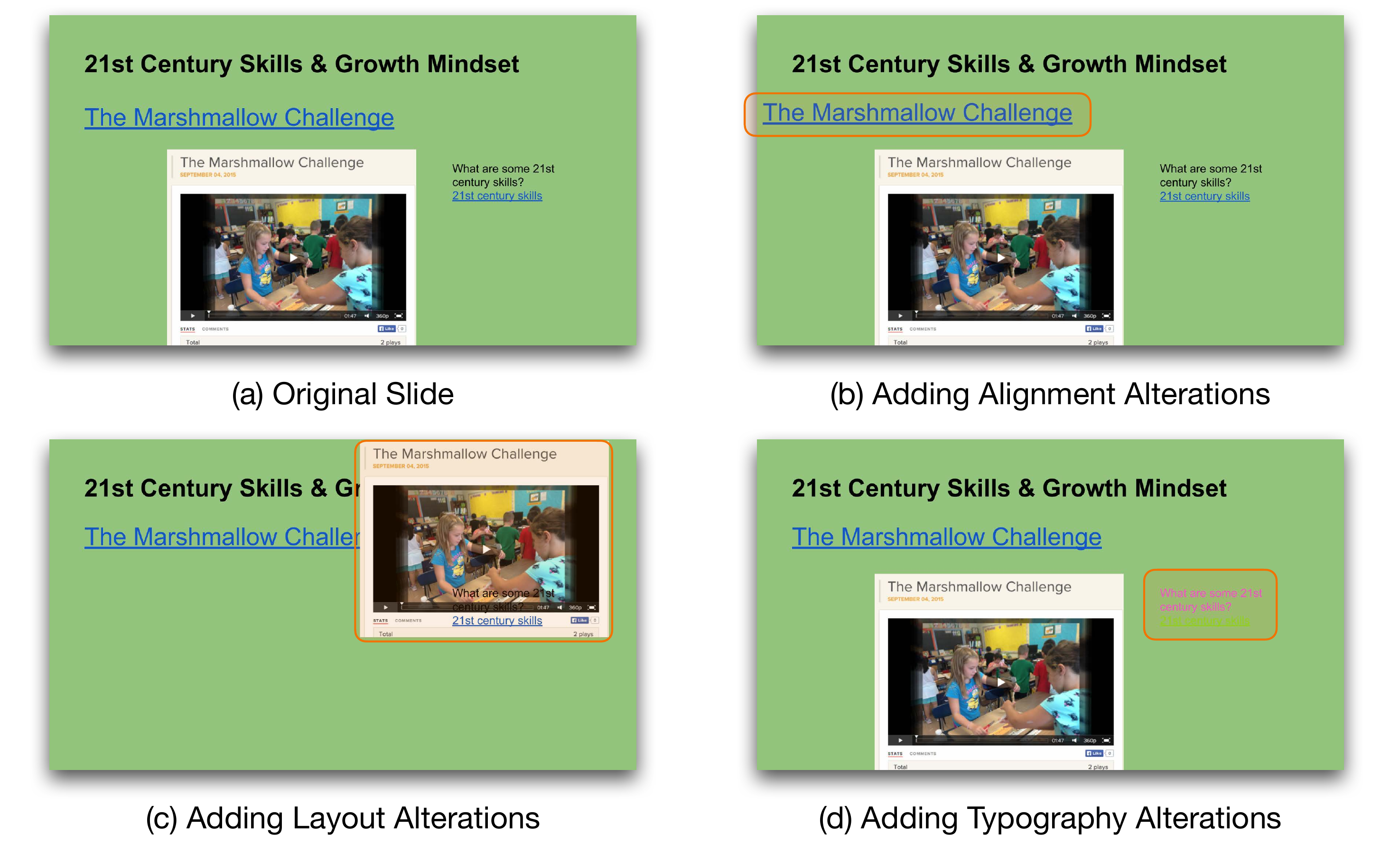}
    \caption{\codename dataset alteration example 3.}
    \label{fig:alt_eg_3}
\end{figure*}

We provide three examples (Figure \ref{fig:alt_eg_1}, \ref{fig:alt_eg_2}, \ref{fig:alt_eg_3}) from the dataset to showcase different alterations.

\section{Prompts}
\label{sec:prompts}

We provide different prompts used in \codename evaluations. Each complete prompt used in our evaluation is a combination of the following prompt parts.

\subsection{Input Formats}
\subsubsection{Image Only}
\begin{lstlisting}

You will be provided with an image of a presentation slide with bounding boxes drawn around various elements. 
    
Analyze this slide for design flaws based on the guidance provided.
If you identify a design flaw, specify its issue (flaw) name, explanation, and the location using normalized coordinates (0-1 scale where 0,0 is top-left and 1,1 is bottom-right).

\end{lstlisting}

\subsubsection{Image with Object Representation}
\begin{lstlisting}

You will be provided with two pieces of information:
1. An image of a presentation slide with bounding boxes drawn around elements and labeled with IDs
2. HTML representation of the slide's objects with their properties and corresponding IDs

When identifying issues, specify its issue name, explanation, and the location using normalized coordinates (0-1 scale where 0,0 is top-left and 1,1 is bottom-right).

\end{lstlisting}

\subsection{Evaluation Prompts}

\subsubsection{Baseline (Zero Shot)}

\begin{lstlisting}
Please examine the provided slide and identify any design flaws or deficiencies that negatively impact its effectiveness.

\end{lstlisting}

\subsubsection{High-level Category Prompting}

\begin{lstlisting}
Please examine the provided slide and identify any design flaws or deficiencies that negatively impact its effectiveness.
Consider all aspects of the slide design including layout, typography, color, visual elements, etc.

IMPORTANT: 
- Never identify font-related issues (size, family, weight, color, etc.) based on missing properties in the HTML representation.
- Always verify these properties visually from the image. The HTML representation is incomplete and should not be used as the source of truth for visual properties.
- Do not force-find issues when there are none. If a slide is well-designed and has no issues, return an empty array.

You MUST provide your response following the output format. Remember, if you cannot find any issues, return an empty array.

\end{lstlisting}

\subsubsection{Full Taxonomy Prompting}

\begin{lstlisting}
Please examine the provided slide and determine if the following specific design issues or flaws are present.
Remember to verify all visual properties from the image, not from the HTML representation.

IMPORTANT: 
- For any font-related properties (size, family, weight, color, etc.), verify them visually from the image.
- Do not identify issues just because of missing properties in the HTML representation.
- Do not overthink and force-find issues when there are none. If this specific issue is not present, set issue_present to false. We do not want you to over-identify issues causing false positives.
- If you cannot find any issues, return an empty array.

Potential Categories:
COMPOSITION & LAYOUT RELATED ISSUES:
  - "Poor Visual Hierarchy" - Elements lack clear importance levels, making it hard to identify the main point. This is not common, because for humans, it is easy to identify the main point. Do not be strict on this one.
  - "Cluttered Layout" - Too many elements crowded together, overwhelming the viewer. Distinguish with Excessive Text Volume, which is about too much text in text boxes. This is for general layout.
  - "Unbalanced Space Distribution or Gapping" - Uneven use of space, with crowded areas alongside empty ones. This is less common, choose with caution. Because some space distribution is intentional.
  - "Object Alignment Issues" - Elements not properly aligned with each other.
  - "Content Overflow/Cut-off" - Text or content extends beyond visible boundaries. This is a common issue. This is when an object is out of the slide boundary, or a text box is cut-off too early into lines that are not supposed to be cut-off.
  - "Occluded Content" - This is very common. Choose when some element is blocking the view of other elements, even it's just a small part of the element.
  
  TYPOGRAPHY ISSUES:
  - "Poor Text Hierarchy" - No clear distinction between headings, subheadings, and body text.  This is not common, because for humans, it is easy to identify the main point. Do not be strict on this one.
  - "Illegible Typeface Selection or Usage" - Font choice is too decorative or complex to read easily.
  - "Improper Font Sizing" - Text size is small somewhere or large somewhere else. Do not choose if the overall text size is small or large. Choose when there is inconsistent text size across the slide.
  - "Excessive Text Volume" - Too much text instead of concise points. Distinguish with Cluttered Layout. This is less common, choose with caution. Because for humans, they are reading the text in their laptop screen.
  - "Improper Text Styling" - Inconsistent or random use of bold, italics, or other formatting. Again, this is only for inconsistent text styling over the slide. If they are consistent overall, even if they are ugly, do not choose this.
  - "Improper Line/Character Spacing" - Text spacing is too tight or too loose. Distinguish with excessive text volume. A text box can have few text lines but still have proper spacing.
  
  COLOR ISSUES:
  - "Insufficient Color Contrast for Readability" - Text and background colors too similar. Not very common.
  - "Excessive or Inconsistent Color Usage" - Too many colors used without clear purpose. 
  - "Inappropriate or Mismatched Color Combinations" - Colors clash or create visual strain. 
  
  IMAGERY & VISUALIZATION ISSUES:
  - "Irrelevant Visual Content" - Images don't support or relate to the content message. Choose with caution because people annotated it usually do not pay attention to visual content.
  - "Poor Image Quality/Editing" - Images are blurry, pixelated, or poorly edited.
  - "Improper Image Sizing" - Images sizes are either too big or too small.
  - "Inconsistent Visual Style Usage" - (This is extremely rare, do not choose this unless you are 100 percent sure) Mixing different visual styles across elements.
  
\end{lstlisting}

\subsubsection{Taxonomy + Computational Augmentations}

\begin{lstlisting}
{...full_taxonomy_prompting}

Additionally, I will provide you with some computational data of this slide for more context. 
  Remember, they are just references, and most times, these contexts are not relevant to the issues you are looking for.

  I'll provide you with 
  (1) a gaze map visualization picture that shows where viewers are most likely to focus their attention when viewing this slide, 
  (2) two lists of proximity-based groupings and similarity-based groupings of the slide's elements, and 
  (3) some color metrics for this slide (colorfulness, text element contrast ratio).
  
  Here is the grouping analysis data. The two lists are groups of close elements' IDs. Each group must have at least 2 elements. An empty list means no groupings are found, which is normal:
  {gaze_data}

  Here is the color data (remember, they are not useful for finding issues, just for your reference):
  {color_data}

\end{lstlisting}

\subsection{Result Comparison}

\begin{lstlisting}
You are tasked with analyzing how AI-generated image categories relate to a set of 19 predefined categories for presentation design issues.

INPUT:
1. PREDEFINED CATEGORIES:
  - There are 19 standard categories for presentation analysis:
  {Same categories from full taxonomy prompting}

2. AI-GENERATED RESULTS:
  - These are issues independently identified by an AI system:
{ai_generated_results}

INSTRUCTIONS:
1. For each of the 19 predefined categories above, determine:
  - Whether any AI-identified issues fit this predefined category
  - Which specific AI-identified issues best match this predefined category

2. For each AI-identified issue, find the BEST matching predefined category.
  If the AI-identified issue does not fit any predefined category, IGNORE it.

3. Create a comprehensive analysis showing how each predefined category relates to the AI-identified issues.

IMPORTANT NOTES:
- If an AI-generated category could potentially fit multiple predefined categories, assign it to the BEST matching one.
- If an AI-generated category does not fit any predefined category, ignore it.
- The "present" field should be true ONLY if there is at least one AI-generated category that fits this predefined category.
- Ensure your mappings are logical and consistent.
\end{lstlisting}

\section{Slide Remediation Examples}
\label{sec:fix_example}
We also include examples from the slide remediation study, including generated fix plans in detailed text for a same slide using different LLM variants. 

\begin{figure*}[t]
    \centering
    \includegraphics[width=\textwidth]{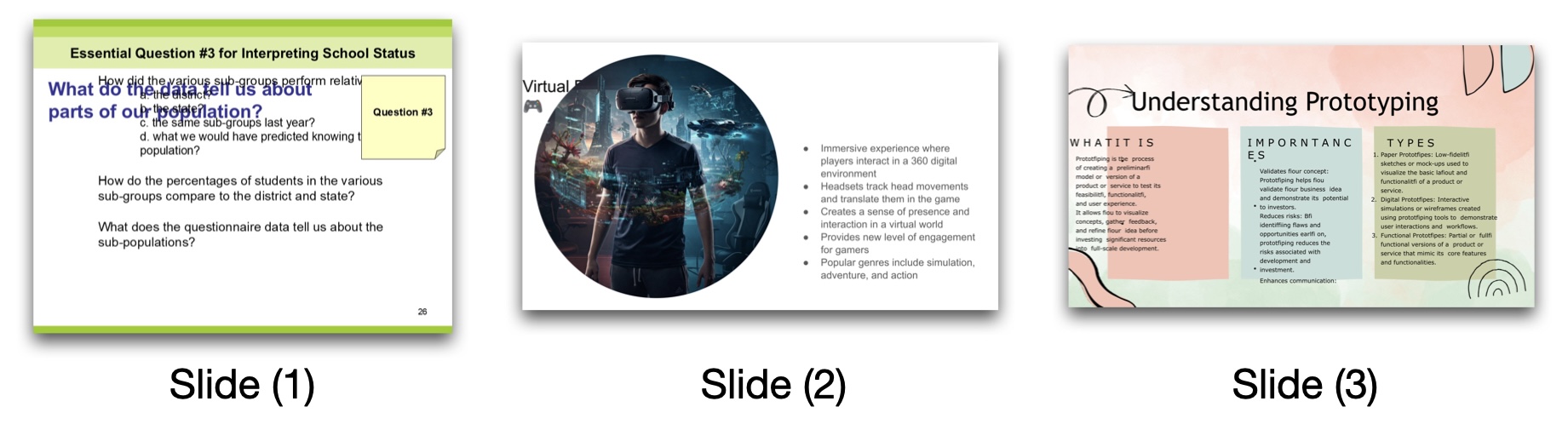}
    \caption{Three example slides from the remediation study.}
    \label{fig:fix_example}
\end{figure*}

\subsection{Example 1}

The original slide is shown in Figure \ref{fig:fix_example}(1).

\begin{lstlisting}
Baseline fix plan:

[
    {
      "category": "Text Overlap",
      "issue": "The main question text overlaps with the subsequent list of questions, making it difficult to read both sections clearly.",
      "fix_plan": "Separate the main question and the list of questions by increasing the vertical space between them. Adjust the positions and ensure that each section has sufficient spacing.",
      "expected_improvements": "Improved readability and clarity by preventing text overlap."
    },
    {
      "category": "Distracting Header Background",
      "issue": "The background color of the header is bright, which might distract readers from the main content.",
      "fix_plan": "Change the background color of the header to a neutral or softer tone that contrasts less with the text.",
      "expected_improvements": "Reduced distraction and better focus on the main content, enhancing overall readability."
    }
]

Taxonomy-based LLM fix plan:

[
    {
      "category": "Cluttered Layout",
      "issue": "The main title overlaps with the body text, making the layout appear crowded and difficult to read.",
      "problem_description": "When text elements overlap, it creates a visually overwhelming and confusing experience for the viewer. It becomes difficult to discern text hierarchy, making it harder to understand the main and supporting points.",
      "fix_plan": "1. Increase the spacing between the main title and body text. \n2. Adjust the size of the main title font if necessary to create more space. \n3. Consider repositioning the main title to the top left corner and reducing its font size slightly.\n4. Ensure there is enough padding around the text elements to avoid overlap.",
      "expected_improvements": "The slide will appear cleaner and more organized. Text hierarchy will be clearer, and viewers will be able to distinguish between the main title and supporting content, improving readability."
    },
    {
      "category": "Occluded Content",
      "issue": "The body text is partially hidden behind the main title, making it difficult to read.",
      "problem_description": "When body text is occluded, it disrupts the flow of information, leading to a poor understanding of the content. Important details may be missed, reducing the effectiveness of the presentation.",
      "fix_plan": "1. Move the body text below the main title to prevent overlap.\n2. Ensure the body text has a clear, distinct area on the slide by allocating sufficient width and height.\n3. If necessary, reduce the font size of the body text or break it into bullet points to enhance clarity.",
      "expected_improvements": "The content will become more accessible and easier to read, ensuring the audience can fully engage with and understand all the information presented."
    }
]
\end{lstlisting}

\subsection{Example 2}

The original slide is shown in Figure \ref{fig:fix_example}(2).

\begin{lstlisting}
Baseline fix plan:

[
    {
      "category": "Text Overlapping with Image",
      "issue": "The text \"Virtual Reality Gaming\" overlaps with the image, making it difficult to read and diminishing the visual appeal of the slide.",
      "fix_plan": "Move the text box containing 'Virtual Reality Gaming' to the right, where it doesn't overlap with the image. Adjust the font size if necessary to maintain balance and readability.",
      "expected_improvements": "This will increase readability and enhance the overall aesthetic of the slide by ensuring all elements are clearly visible and properly aligned."
    },
    {
      "category": "Text Alignment and Size",
      "issue": "The main text box containing bullet points starts too far to the right, leaving excessive space between the image and text.",
      "fix_plan": "Shift the main text box to the left, closer to the image. Ensure there is consistent padding between the text and the image to achieve a balanced look. Adjust the text box width to ensure all content fits properly.",
      "expected_improvements": "The slide will appear more balanced, with improved flow between the image and text. This will make the content more engaging and easier to follow."
    }
]

Taxonomy-based LLM fix plan:

[
    {
      "category": "Content Overflow/Cut-off",
      "issue": "The text 'Virtual Reality Gaming' is partially cut off by the left edge of the slide.",
      "problem_description": "When text is cut off, it makes the content difficult to read and unprofessional, negatively impacting the slide's effectiveness and readability.",
      "fix_plan": [
        "1. Select the text box containing 'Virtual Reality Gaming'.",
        "2. Move the text box to the right by adjusting the x-position from 0.0 to a small positive value (e.g., 0.05) to ensure it is fully visible.",
        "3. Ensure the text box width or font size is adjusted if necessary to maintain readability."
      ],
      "expected_improvements": "The text will be fully visible, enhancing readability and professionalism."
    },
    {
      "category": "Occluded Content",
      "issue": "The image is overlapping and partially occludes the text 'Virtual Reality Gaming'.",
      "problem_description": "Overlapping elements can make text unreadable and cluttered, which detracts from the aesthetic appeal and clarity of the slide.",
      "fix_plan": [
        "1. Select the image element.",
        "2. Adjust the x-position of the image from 0.025 to a larger value (e.g., 0.1) to move it slightly to the right.",
        "3. Re-check the position of the text box to ensure no overlap occurs. If overlap still happens, adjust either the image or text box positions further.",
        "4. Optionally, resize the image if space is limited, ensuring it does not compromise the overall layout."
      ],
      "expected_improvements": "The text will no longer be occluded by the image, improving readability and the slide's overall visual balance."
    }
]
\end{lstlisting}

\subsection{Example 3}

The original slide is shown in Figure \ref{fig:fix_example}(3).

\begin{lstlisting}
Baseline fix plan:

[
    {
      "category": "Spelling Errors",
      "issue": "The text contains misspellings such as 'Prototfiping' instead of 'Prototyping' and 'fiour' instead of 'your'.",
      "fix_plan": "Review and correct all text for spelling errors. Use a spell checker or proofreading tool to ensure accuracy.",
      "expected_improvements": "Improved readability and professionalism of the slide, resulting in clearer communication of key points."
    },
    {
      "category": "Empty Text Boxes",
      "issue": "Some text boxes are empty, indicating possible missing content or placeholders.",
      "fix_plan": "Identify the purpose of each text box and fill them with appropriate content or remove those that are not needed.",
      "expected_improvements": "A complete and polished slide presentation with no unintentional gaps in content."
    },
    {
      "category": "Floating Dots or Small Elements",
      "issue": "There are very small image elements resembling dots that may distract viewers.",
      "fix_plan": "Remove or consolidate these small elements to maintain focus on essential content.",
      "expected_improvements": "A cleaner and more aesthetically pleasing slide without unnecessary distractions."
    },
    {
      "category": "Misalignment of Text Blocks",
      "issue": "Text blocks are not well aligned, impacting readability and aesthetics.",
      "fix_plan": "Align all text boxes uniformly. Use alignment guides or a grid to ensure consistent spacing.",
      "expected_improvements": "Enhanced visual balance and easier navigation through the slide's content."
    }
]

Taxonomy-based LLM fix plan:

[
    {
      "category": "Object Alignment Issues",
      "issue": "Text headings are not properly aligned vertically with their associated text boxes.",
      "problem_description": "Misaligned headings create a disorganized appearance and make it difficult for viewers to navigate and comprehend the content smoothly.",
      "fix_plan": [
        "Open the slide in a presentation editing tool.",
        "Select the text heading 'WHAT IT IS'.",
        "Align the heading centrally at the top of the associated text box.",
        "Repeat for 'IMPORTANCES' and 'TYPES', ensuring vertical alignment with their corresponding text boxes.",
        "Use consistent spacing and alignment guides if available in the software."
      ],
      "expected_improvements": "Aligning headings will create a more coherent and professional appearance, improving readability and making it easier for the audience to follow the information."
    },
    {
      "category": "Content Overflow/Cut-off",
      "issue": "Text content 'IMPORTANCES' is cut off within the text box.",
      "problem_description": "Cut-off text can cause confusion and prevent the message from being fully conveyed, reducing the effectiveness of the presentation.",
      "fix_plan": [
        "Select the text box containing 'IMPORTANCES'.",
        "Resize the text box to fit all the content; increase the height if necessary.",
        "Alternatively, adjust the font size or line spacing to ensure all content is visible without overcrowding the text box."
      ],
      "expected_improvements": "Ensuring all text is visible will enhance understanding and maintain audience engagement by allowing the full message to be presented clearly."
    },
    {
      "category": "Occluded Content",
      "issue": "Decorative elements overlap with the text heading 'Understanding Prototyping'.",
      "problem_description": "Overlapping elements can distract the audience and make the heading difficult to read, detracting from the overall visual impact.",
      "fix_plan": [
        "Select the decorative elements (arrows).",
        "Move or resize these elements so they do not interfere with the text.",
        "Ensure sufficient space is left around the heading for clear visibility.",
        "Consider changing the layering of elements if necessary, sending decorative elements to the back."
      ],
      "expected_improvements": "Removing the overlap will improve the slide's aesthetic and readability, ensuring the main heading is clear and attention-grabbing."
    }
]
\end{lstlisting}

\end{document}